\documentclass[11pt]{article}
\usepackage{jcappub,natbib}
\input{colordvi.tex}
\title{Signatures of anisotropic sources in the squeezed-limit
bispectrum of the cosmic microwave background}
\author[a]{Maresuke Shiraishi,}
\author[b,c,d]{Eiichiro Komatsu,}
\author[e,f]{Marco Peloso,}
\author[g]{and Neil Barnaby}
\affiliation[a]{Department of Physics and Astrophysics,
Nagoya University, Nagoya, Aichi 464-8602, Japan}
\affiliation[b]{Max-Planck-Institut f\"{u}r Astrophysik,
Karl-Schwarzschild Str. 1, 85741 Garching, Germany}
\affiliation[c]{Kavli Institute for the Physics and
Mathematics of the Universe, Todai Institutes for Advanced Study, the
University of Tokyo, Kashiwa, Japan 277-8583 (Kavli IPMU, WPI)}
\affiliation[d]{Texas Cosmology Center and the Department of Astronomy,
The University of Texas at Austin, 1 University Station, C1400, Austin,
TX 78712, USA}
\affiliation[e]{School of Physics and Astronomy, University of
Minnesota, Minneapolis 55455, USA}
\affiliation[f]{INFN, Sezione di Padova, via Marzolo 8, I-35131, Padova, Italy}
\affiliation[g]{Department of Applied Mathematics and Theoretical
Physics, Center for Mathematical Sciences, Wilberforce Road, Cambridge,
CB3 0WA, UK}

\def\ga{\mathrel{\raise.3ex\hbox{$>$\kern-.75em\lower1ex\hbox{$\sim$}}}}
\def\la{\mathrel{\raise.3ex\hbox{$<$\kern-.75em\lower1ex\hbox{$\sim$}}}}

\abstract{%
The bispectrum of primordial curvature perturbations in the squeezed
configuration, in which one wavenumber, 
$k_3$, is much smaller than the other two, $k_3\ll k_1\approx k_2$,
plays a special role in constraining the 
physics of inflation. In this paper we study a new phenomenological
signature in the squeezed-limit bispectrum: namely, the amplitude of the
squeezed-limit bispectrum depends on an angle between ${\bf k}_1$ and
${\bf k}_3$ such that $B_\zeta(k_1, k_2, k_3) \to 2 \sum_L c_L
P_L(\hat{\bf k}_1 \cdot \hat{\bf k}_3) P_\zeta(k_1)P_\zeta(k_3)$,
where $P_L$ are the Legendre polynomials.
While $c_0$ is related to the usual 
local-form $f_{\rm NL}$ parameter as $c_0=6f_{\rm NL}/5$, the
higher-multipole coefficients, $c_1$, $c_2$, etc., have not been
constrained by the data. Primordial curvature perturbations sourced by
large-scale magnetic fields generate non-vanishing $c_0$, $c_1$, and
$c_2$. Inflation models whose action contains a term like $I(\phi)^2 F^2$
generate $c_2=c_0/2$. A recently proposed ``solid inflation'' model
generates $c_2\gg c_0$. 
A cosmic-variance-limited experiment measuring temperature anisotropy of
the cosmic microwave background up to $\ell_{\rm max}=2000$ is able to
measure these coefficients down to $\delta c_0=4.4$, $\delta c_1=61$,
and $\delta c_2=13$ (68\%~CL). We also find that $c_0$ and $c_1$, and
$c_0$ and $c_2$, are nearly uncorrelated. Measurements of these
coefficients will open up a new window into the physics of inflation
such as the existence of vector fields during inflation or non-trivial
symmetry structure of inflaton fields. Finally, we show that the
original form of the Suyama-Yamaguchi inequality does not apply to the
case involving higher-spin fields, but a generalized form does.
}

\begin{document}

\begin{flushright}  UMN-TH-3135/13 \end{flushright}

\maketitle
\flushbottom

\section{Introduction}
Convincing detection of the so-called ``local-form'' three-point
correlation function (bispectrum) of primordial curvature perturbations
from inflation would rule out all single-field inflation
models \cite{creminelli/zaldarriaga:2004,komatsu/etal:prep}, provided
that an initial quantum 
state of the curvature perturbation is in a preferred state called
the Bunch-Davies state \cite{agullo/parker:2011, ganc:2011, Chialva:2011hc} and that the
curvature perturbation does not evolve outside the horizon due to a
non-attractor solution
\cite{namjoo/firouzjahi/sasaki:prep,chen/etal:prep}.\footnote{Also see workshop
summaries of ``Critical Tests of Inflation Using Non-Gaussianity'' in {\sf
http://www.mpa-garching.mpg.de/\textasciitilde{}komatsu/meetings/ng2012/}.} 

The curvature perturbation, $\zeta$, is defined as a trace part of 
space-space components of the metric perturbation, $\delta
g_{ij}=a^{2}(t)e^{2\zeta}\delta_{ij}$, in a uniform density gauge. The
bispectrum of $\zeta$ is defined as
$\langle\zeta_{{\bf k}_1} \zeta_{{\bf k}_2} \zeta_{{\bf k}_3} \rangle
= (2\pi)^3 \delta^{(3)}\left({{\bf k}_1} + {{\bf k}_2} + {{\bf k}_3}\right)
B_\zeta (k_1,k_2,k_3)$,  
and the local-form bispectrum is defined as
$B_\zeta (k_1,k_2,k_3)=\frac65f_{\rm
NL}\left[P_\zeta(k_1)P_\zeta(k_2)+P_\zeta(k_2)P_\zeta(k_3)+P_\zeta(k_3)P_\zeta(k_1)\right]$
(e.g., \cite{Komatsu:2001rj}), where $P_\zeta(k)\propto k^{n_s-4}$ is the power
spectrum of the primordial curvature perturbation with $n_s=0.96\pm
0.01$ \cite{hinshaw/etal:prep,hou/etal:prep,sievers/etal:prep}. This
means that the local-form bispectrum has the largest amplitude in the
so-called squeezed configuration, in which the smallest wavenumber,
$k_3$, is much smaller than the other two, i.e., $k_3\ll k_1\approx
k_2$ \cite{babich/creminelli/zaldarriaga:2004}. In this limit the
local-form bispectrum is given by $B_\zeta\to 
\frac{12}{5}f_{\rm NL}P_\zeta(k_1)P_\zeta(k_3)$, and all attractor single-field inflation
models with a Bunch-Davies initial state give $f_{\rm
NL}=\frac5{12}(1-n_s)$ \cite{maldacena:2003,creminelli/zaldarriaga:2004}.

The current best limit on $f_{\rm NL}$ is $f_{\rm NL}=37\pm
20$~(68\%~CL), which was obtained from the {\sl Wilkinson Microwave
Anisotropy Probe} ({\sl 
WMAP}) 9-year data with the expected ISW-lensing bias removed
\cite{bennett/etal:prep}. The forthcoming {\sl Planck} data are expected
to reduce the error bar by a factor of four \cite{Komatsu:2001rj}.

If the {\sl Planck} collaboration finds  evidence for
$f_{\rm NL}$, or the lack thereof, what is next? Measuring the 
local-form four-point function 
(trispectrum)
\cite{okamoto/hu:2002,kogo/komatsu:2006,boubekeur/lyth:2006} to check
the 
so-called Suyama-Yamaguchi inequality between the amplitude of the
local-form trispectrum and $f_{\rm NL}$, i.e., $\tau_{\rm NL}\ge (6f_{\rm NL}/5)^2$
\cite{suyama/yamaguchi:2007,komatsu:2010,sugiyama/komatsu/futamase:2011,smith/loverde/zaldarriaga:2011,Lewis:2011au,Bramante:2011zr,sugiyama:2012,assassi/baumann/green:2012},
would be an important next step to understand the nature
of sources of non-Gaussianity (or the absence thereof). We shall discuss
the Suyama-Yamaguchi inequality within the context of higher-spin fields
in section~\ref{sec:SY}.

\begin{figure}[t]
  \begin{tabular}{cc}
    \begin{minipage}{0.5\hsize}
  \begin{center}
    \includegraphics[width = 7.5 cm]{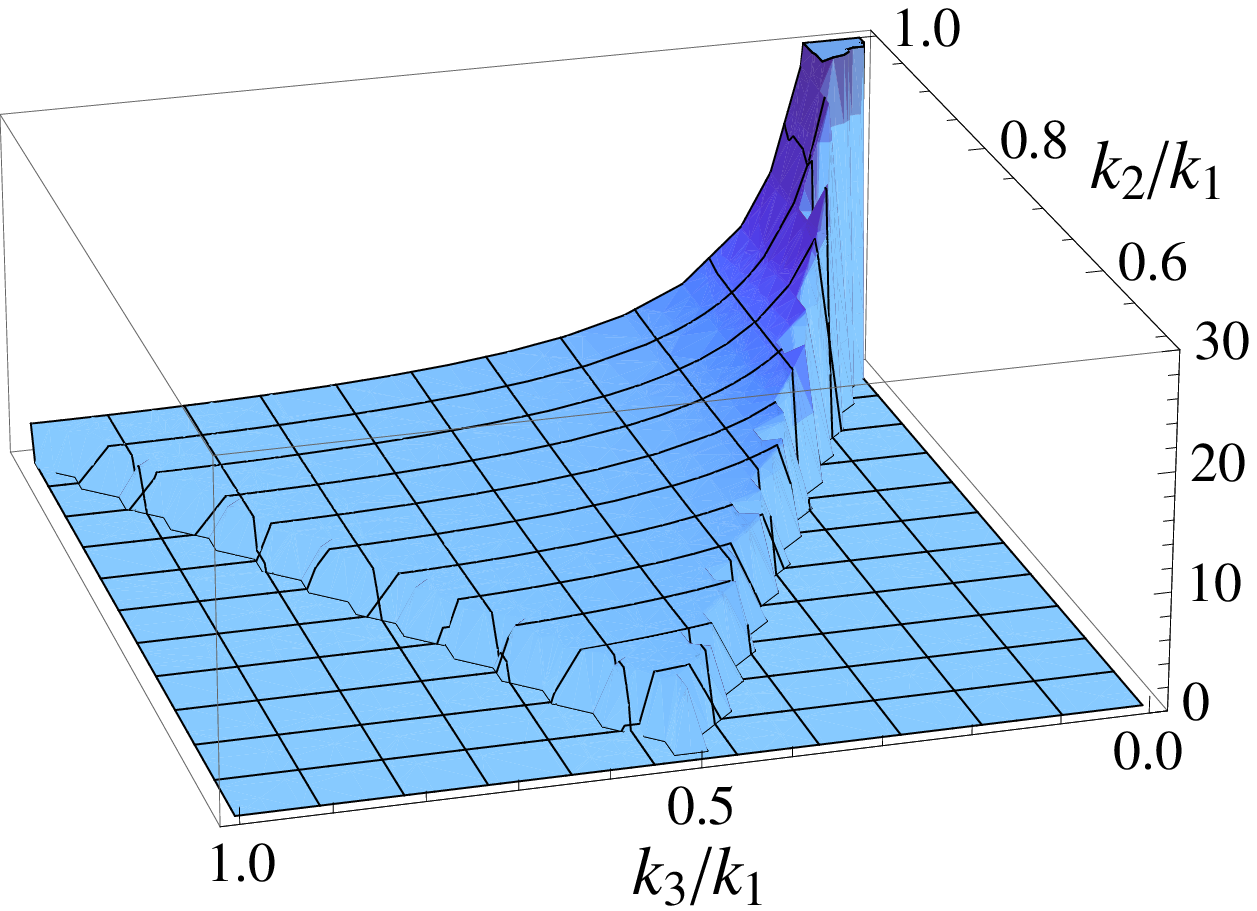}
  \end{center}
\end{minipage}
\begin{minipage}{0.5\hsize}
  \begin{center}
    \includegraphics[width = 7.5 cm]{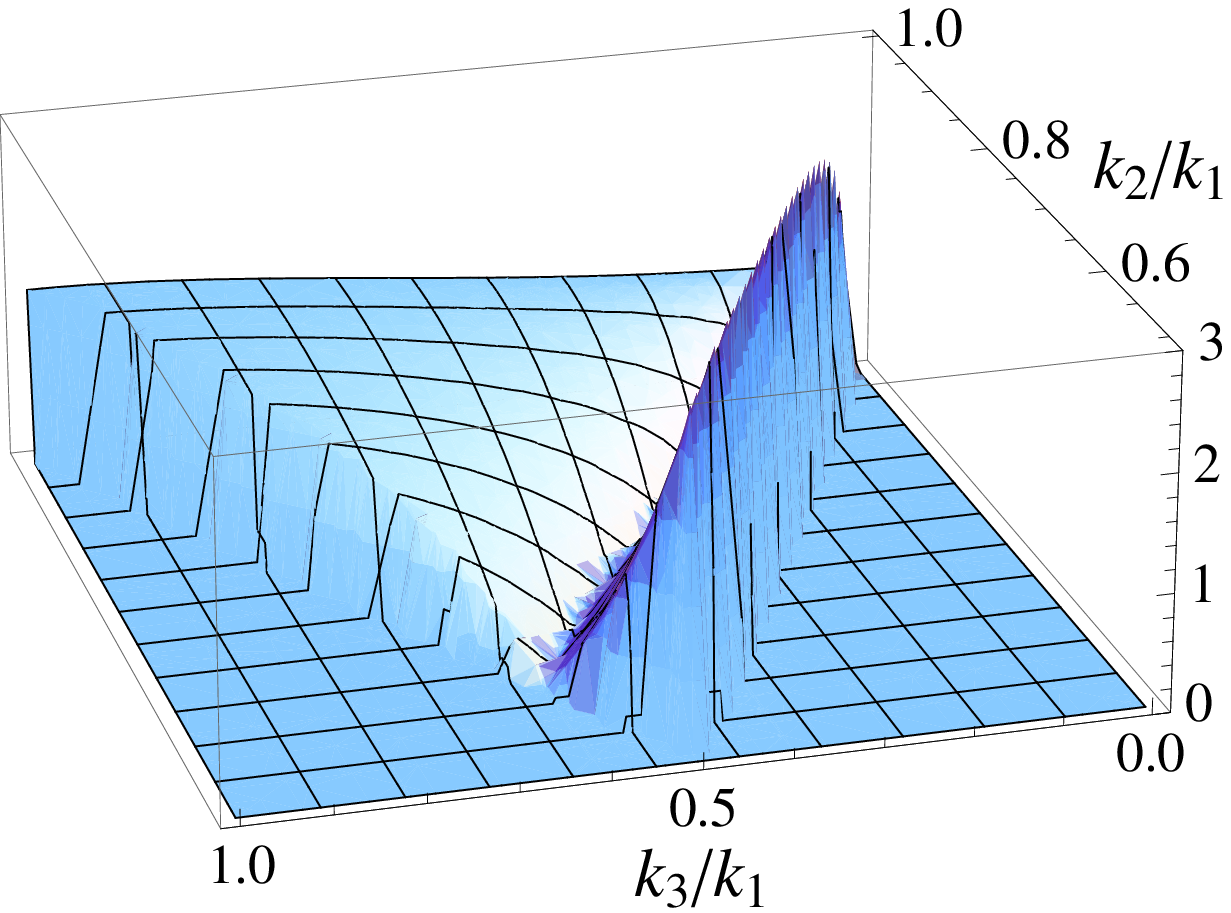}
  \end{center}
\end{minipage}
\end{tabular}
\\
  \begin{tabular}{c}
    \begin{minipage}{1.0\hsize}
  \begin{center}
    \includegraphics[width = 7.5 cm]{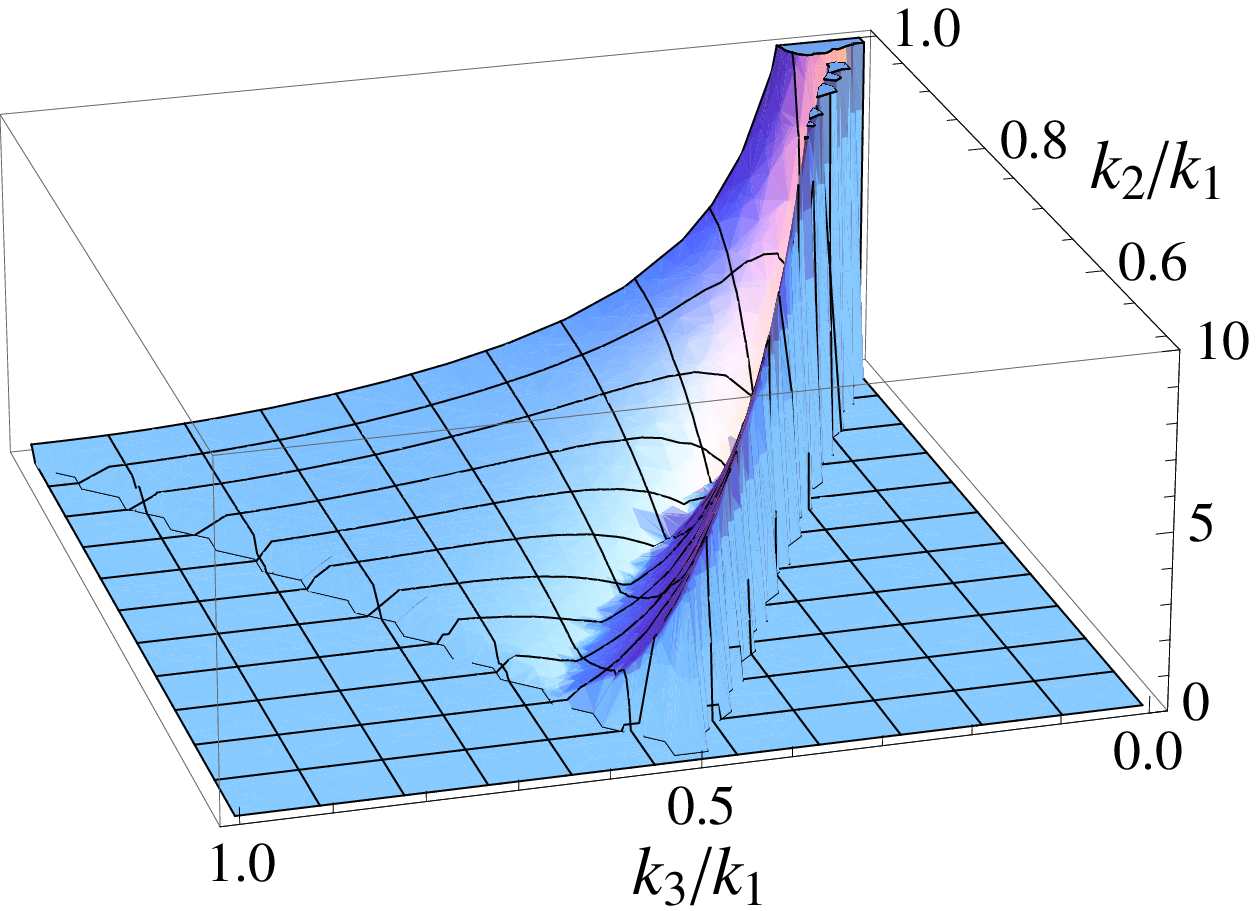}
  \end{center}
\end{minipage}
\end{tabular}
  \caption{Absolute values of the shape function of $L=0$, $(k_1 k_2 k_3)^2 S_0$ (top left panel), that of $L=1$, $(k_1 k_2 k_3)^2 S_1$ (top right panel), and that of $L=2$, $(k_1 k_2 k_3)^2 S_2$ (bottom panel). 
We restrict the plot range to $k_3 \leq k_2 \leq k_1$
 and $|k_1 - k_2| \leq k_3 \leq k_1 + k_2$ for symmetry and the triangular
 condition. The shape of $L=2$ peaks at the squeezed configuration,
 $k_3/k_1\ll 1$ and $k_2/k_1\approx 1$, in the same way as that of $L=0$ whereas the shape of $L=1$ is suppressed at the squeezed configuration. While the shape function of $L=0$ has positive values for all $k_2 / k_1$ and $k_3 / k_1$, those of $L = 1$ and $2$ have negative values except in the flattened configurations, $k_2/k_1 + k_3/k_1 \approx 1$.}
\label{fig:RRR} 
\end{figure}

Can we learn more about sources of non-Gaussianity by further
scrutinizing the behavior of the bispectrum in the squeezed configuration? The
answer is yes, and this is the main goal of this paper. Namely, in this
paper, we shall investigate phenomenological consequences of the 
following new parametrization of the bispectrum of primordial curvature
perturbations:
\begin{equation}
B_\zeta (k_1,k_2,k_3) 
= \sum_L c_L P_L(\hat{\bf k}_1 \cdot \hat{\bf
 k}_2)P_\zeta(k_1)P_\zeta(k_2)+(\mbox{2 perm})~, 
\label{eq:bis_curv_general} 
\end{equation}
where $P_L(\mu)$ is the usual Legendre polynomials, i.e., 
$P_0(\mu)=1$, $P_1(\mu)=\mu$, and $P_2(\mu)=\frac12(3\mu^2-1)$. 
Here, $c_0$ is equal to $6f_{\rm NL}/5$.\footnote{Note that, due to
symmetry, the $c_1$ term as well as any odd $L$ 
terms vanish in the exact squeezed limit, i.e., 
$\lim_{k_3\rightarrow 0}B_\zeta(k_1,k_2,k_3)=
2[c_0+c_2P_2(\mu_{13})+c_4P_4(\mu_{13})+\dots]P_\zeta(k_1)P_\zeta(k_3)$, where
$\mu_{13}\equiv \hat{\bf k}_1 \cdot \hat{\bf
 k}_3$ (also see ref.~\cite{Lewis:2011au}).
As a result, when we analyze the CMB data, the error bar of
 $c_1$ is much bigger than the error bars of $c_0$ and $c_2$, as we
 shall show in section~\ref{sec:2DFisher}. On the other hand, the error bars
 of $c_0$  and $c_2$ are expected to be comparable: we shall show that
 they are related by $\delta c_2\approx 3\delta c_0$ in section~\ref{sec:2DFisher}.}

Why consider $c_L$ with $L\ge 1$? These coefficients appear to be
sensitive to the existence of vector fields. For example, curvature
perturbations sourced by primordial magnetic fields produce non-zero $c_1$ and $c_2$ \cite{shiraishi/etal:2012, shiraishi:2012}. 
Curvature perturbations sourced by a $\frac14I(\phi)^2F^2$ term in Lagrangian
produce $c_2=c_0/2$ \cite{Barnaby:2012tk, Bartolo:2012sd}.
These coefficients are also sensitive to the existence of a non-trivial
realization of SO(3) rotational symmetry during inflation: a recently
proposed ``solid inflation'' model 
produces $c_2\gg c_0$ \cite{Endlich:2012pz}. 
While the second-order effects in General Relativity also induce
non-trivial angular dependence in the bispectrum, it disappears in the
squeezed limit \cite{liguori/etal:2006}, and thus will not be considered
in this paper.\footnote{A correlation between the integrated
Sachs--Wolfe effect 
and the gravitational lensing of CMB produces an angle-dependent squeezed-limit
bispectrum of the CMB temperature anisotropy
\cite{goldberg/spergel:1999}, which goes as 
$\cos(2\phi)$ in the flat-sky approximation where
$\cos\phi\equiv\hat{\boldsymbol{\ell}_1}\cdot 
\hat{\boldsymbol{\ell}_3}$
\cite{boubekeur/etal:2009,lewis/challinor/hanson:2011}. We
ignore this secondary effect in this paper.}

While we assume that the coefficients, $c_L$, do not depend on
wavenumbers, it is entirely possible that they do. There are various
ways in which $c_0=6f_{\rm NL}/5$ depends 
on wavenumbers
\cite{byrnes/etal:2010a,byrnes/etal:2010b,shandera/dalal/huterer:2011,byrnes/gong:2013,agullo/parker:2011,ganc:2011,ganc/komatsu:2012,agullo/shandera:2012}. Particularly
interesting possibilities are strongly infrared-divergent $c_1$ and
$c_2$, which can naturally give rise to dipoler (i.e., hemispherical)
and quadrupolar modulations, respectively, of the observed power
spectrum in our sky 
\cite{schmidt/hui:2013}.  

In section~\ref{sec:motivation},
we shall briefly review these three scenarios to motivate our choice of
parametrization given in eq.~(\ref{eq:bis_curv_general}). Specifically, we
review non-Gaussianities 
generated from: (1) large-scale magnetic fields after
inflation in
section~\ref{subsec:PMF}; (2) a vector field coupled to the inflaton field,
$\phi$, through a dilaton-like coupling $I^2(\phi)F^2$ in
section~\ref{subsec:vectors}; and 
(3) solid inflation, in which the inflaton field is a part of a vector
multiplet, $\left\{ \phi^1, \phi^2, \phi^3 \right\}$, in section~\ref{subsec:solid}.

What do the shapes of $L=1$ and $L=2$ terms look like? Using the
triangular condition of three wavevectors, ${{\bf k}}_1 + {{\bf k}}_2 +
{{\bf k}}_3 = 0$, the cosine between wavenumbers can be written as,
e.g., $\hat{\bf k}_1\cdot\hat{\bf
k}_2=(k_3^2-k_1^2-k_2^2)/(2k_1k_2)$. Then, for a scale-invariant
power spectrum of $\zeta$, $P_\zeta(k) = \frac{2 \pi^2}{k^3} A_S$, 
eq.~(\ref{eq:bis_curv_general}) can be re-written as
\begin{eqnarray}
B_\zeta(k_1,k_2,k_3) = (2 \pi^2 A_S)^2  \sum_L c_LS_L(k_1, k_2, k_3) ~, 
\end{eqnarray}
where
\begin{eqnarray}
S_{0}(k_1, k_2, k_3) 
&=& \left( \frac{1}{k_1^3 k_2^3} + \mbox{2 perm} \right) ~, \\ 
S_{1}(k_1, k_2, k_3) 
&=& \left( \frac{k_1^2}{2 k_2^4 k_3^4} + \mbox{2 perm} \right) 
- \left( \frac{1}{2 k_1^4 k_2^2} + \mbox{5 perm} \right)~, \\ 
S_{2}(k_1, k_2, k_3)  
&=& \left( \frac{3 k_1^4}{8 k_2^5 k_3^5} + \mbox{2 perm} \right) 
- \left( \frac{3 k_1^2}{4 k_2^5 k_3^3} + \mbox{5 perm} \right) \nonumber \\ 
&&+ \left( \frac{3}{8 k_1^5 k_2} + \mbox{5 perm} \right) 
+ \left( \frac{1}{4 k_1^3 k_2^3} + \mbox{2 perm} \right) ~.
\end{eqnarray}

In figure~\ref{fig:RRR}, we show the shape functions $S_0$ (top left panel), $S_1$ (top right one) and $S_2$ (bottom one). The bottom panel shows that the bispectrum for $L=2$ peaks in the squeezed limit ($k_3 \ll k_1 \approx k_2$) in the same way as that for $L=0$, whereas
the top right panel shows that the bispectrum for $L=1$ is suppressed in the
squeezed limit (also see footnote 2). The bispectrum for $L=1$ peaks when $k_2 + k_3 = k_1$. This signature may also be seen in the ``flattened bispectrum template'' defined by eq.~(5.1) of ref.~\cite{Meerburg:2009ys}. Nevertheless, the correlation coefficient between these bispectra is $0.196$~\footnote{We thank  Christian Byrnes for suggesting to compute this correlation.}; these are only weakly correlated because the flattened template does not change sign while $S_1$ is positive in the flattened configurations and is negative otherwise.

The rest of this paper is organized as follows. 
In section~\ref{sec:CMB_bis}, we derive both the full-sky and flat-sky
formulae of the bispectrum of temperature anisotropies of the cosmic
microwave background (CMB) induced by the angle-dependent bispectrum
given in eq.~(\ref{eq:bis_curv_general}), and analyze their
behaviors.
We also estimate the error bars of $c_L$
for $L=0$, 1, and 2, expected for a cosmic-variance-limited CMB
experiment measuring temperature anisotropy up to $\ell_{\rm max}=2000$.
In section~\ref{sec:SY}, we revisit the Suyama-Yamaguchi inequality within
the context of higher-spin fields such as those discussed in this paper.
 We conclude in section~\ref{sec:conclusion}.
In appendix~\ref{app:flat}, we discuss the precision of the flat-sky
approximation. In appendix~\ref{appen:SW}, we derive the CMB
bispectrum in the 
Sachs--Wolfe limit (in which the temperature anisotropy is given by
$\delta T/T=-\zeta/5$). In appendix~\ref{appen:3Dfit}, we present the
full Fisher matrix for $c_0$, $c_1$, and $c_2$.

Throughout this paper, we adopt the following convention for Fourier
transformation of an arbitrary function, $f({\bf x})$: 
$f({\bf x}) = \int \frac{d^3 {\bf k}}{(2\pi)^3}f({\bf k}) e^{i {\bf k}
 \cdot {\bf x}}$. 

\section{Theoretical motivation}
\label{sec:motivation}
\subsection{Helical and non-helical magnetic fields} \label{subsec:PMF}

Astrophysical observations suggest the existence of  magnetic
fields on the order of $10^{-6}$~G in galaxies and cluster of galaxies
\cite{Kronberg:2007dy, 
Bernet:2008, Wolfe:2008, Fletcher:2010}. 
There is also indirect evidence for the existence of  magnetic fields on
the order of $10^{-20} - 10^{-14}$~G in the inter-galactic medium (IGM)
\cite{Neronov:1900zz, Dolag:2010,Tavecchio:2010, Takahashi:2011}.

There is yet no compelling model for how these vector fields can be
generated during inflation, as the existing 
models suffer from strong backreaction or strong coupling problems 
\cite{Barnaby:2012tk,Kanno:2009ei,Demozzi:2009fu,Demozzi:2012,Suyama:2012,Fujita:2012rb}  
(see the more detailed discussion in the next subsection).
\footnote{See e.g., refs.~\cite{Koivisto:2011rm, Urban:2012ib} for attempts to avoid such problems.} 
Here we
simply assume that a magnetic field has been
generated,\footnote{While 
we use the term ``magnetic field'' and ``electric field'' here and in
the next subsection, these fields are not necessarily the usual
electromagnetic fields. For the discussion in this subsection, it is
sufficient to have some vector field whose anisotropic stress decays as
$T^i_j-\frac13\delta^i_jT^k_k\propto a^{-4}$ on
super-horizon scales. On the other hand, the anisotropic 
stress on super-horizon scales is constant during inflation
(disregarding slow-roll corrections) for the case we discuss in the
next subsection. The important feature of these models is
that the anisotropic stress scales with $a$ in the same way as the
isotropic pressure dominating the universe: for the former case, it
scales in the same way the radiation pressure does, and for the latter
it scales in the same way the inflaton pressure does.}
and study its 
impact on the primordial perturbations during the radiation 
era and recombination. In the next subsection, we discuss the additional
signatures that take place if vector fields are coupled to the
inflaton field.

A large amount of literature exist in the studies of 
effects of vector fields on CMB anisotropies and the large-scale
structure of the universe. See, e.g., refs.~\cite{Durrer:1998ya, Mack:2001gc, Lewis:2004ef, Yamazaki:2008gr,Paoletti:2008ck, Finelli:2008xh, Shaw:2009nf, Shaw:2010ea,
Paoletti:2010rx, Paoletti:2012bb, Yadav:2012uz,shiraishi/saga/yokoyama:2012,kunze:2013,kunze:prep} for effects on the two-point
correlation functions, and refs.~\cite{Seshadri:2009, Caprini:2009, Trivedi:2010, Cai:2010, shiraishi/etal:2010b, shiraishi/etal:2011b,
shiraishi/etal:2011c, shiraishi/etal:2012, shiraishi:2012} for those on
higher-order correlation functions.

Let us assume that super-horizon vector perturbations were
produced during inflation, and they generated large-scale magnetic fields.
 The anisotropic stress of this magnetic field sources the growth of
 curvature perturbation via Einstein's field equations during the
 radiation era. However, after the decoupling of neutrinos at a few MeV,
 the magnetic anisotropic stress is compensated by the neutrino
 anisotropic stress, and the curvature perturbation on super-horizon
 scales becomes a constant. This constant curvature
 perturbation survives till the recombination epoch, and seeds
 additional CMB anisotropies. The solution of  curvature perturbations
 on super-horizon scales is determined by the traceless projection of the
 magnetic anisotropic stress as \cite{Shaw:2009nf}  
\begin{eqnarray}
\zeta_{\bf k} \approx 0.9 \ln \left( \frac{\tau_\nu}{\tau_B} \right) 
 \left( \hat{k}^j \hat{k}_i - \frac{1}{3} \delta^j_{~i} \right) 
\frac{1}{4\pi \rho_{\gamma, 0}} \int \frac{d^3 {\bf k'}}{(2\pi)^3} 
B^i({\bf k'}) B_j({\bf k} - {\bf k'}),
\end{eqnarray}
where $\tau_B$ and $\tau_\nu$ denote the conformal time of the
generation of magnetic fields  and that of the decoupling of
neutrinos, respectively, and $\rho_{\gamma, 0}$ is the present-day value of
the photon energy density. This equation shows that, even under the
assumption that the magnetic field itself is a Gaussian variable,
the curvature perturbations become highly non-Gaussian. 

Angular dependence of the power spectrum and bispectrum
arises due to the spin-1 nature of magnetic fields. The magnetic field
vector is transverse, 
$k^iB_i=0$, and thus it is expanded using the spin-1 polarization
vector, $e_i^{(\sigma)}({\hat{\mathbf k}})$: $B_i({\mathbf x})=(2\pi)^{-3}\int
d^3{\mathbf k}~B_i(\mathbf{k})e^{i{\mathbf k}\cdot{\mathbf x}}=(2\pi)^{-3}\int
d^3{\mathbf k}~\sum_{\sigma=\pm 1} B^{\sigma}(\mathbf{k})e_i^{(\sigma)}({\hat{\mathbf
k}})e^{i{\mathbf k}\cdot{\mathbf x}}$, where
$\sigma$ denotes two circular polarization states. Then, the power
spectrum of magnetic fields can be decomposed into the ``non-helical,''
$P_B(k)$, and ``helical,'' $P_{\cal B}(k)$, components as
\cite{caprini/durrer/kahniashvili:2003} 
\begin{eqnarray}
\nonumber
\left\langle B_i ({\bf k}) B_j ({\bf k'}) \right\rangle 
&=& 
- \frac{(2 \pi)^3}{2} \delta^{(3)}({\bf k} + {\bf k'})
\sum_{\sigma=\pm 1}
 \left[ P_B(k) -\sigma P_{{\cal B}}(k) \right]e_i^{(\sigma)}({\hat{\mathbf
 k}})e_j^{(-\sigma)}({\hat{\mathbf k}}) \\
&=&\frac{(2 \pi)^3}{2}\delta^{(3)}({\bf k} + {\bf k'})
 \left[ (\delta_{ij} - \hat{k}_i \hat{k}_j) P_B(k) 
+ i \epsilon_{ijl} \hat{k}^l P_{{\cal B}}(k) \right],
\label{eq:mag_power} 
\end{eqnarray} 
where $\epsilon_{ijl}$ is the antisymmetric tensor normalized as
$\epsilon_{123} = 1$. These power spectra are defined as 
\begin{eqnarray}
- \langle B^{+}({\mathbf k})B^{+}({\mathbf k}')\rangle
- \langle B^{-}({\mathbf k})B^{-}({\mathbf k}')\rangle 
&=& (2\pi)^3 P_B(k)\delta^{(3)}({\bf k} + {\bf k'}),\\
\langle B^{+}({\mathbf k})B^{+}({\mathbf k}')\rangle
- \langle B^{-}({\mathbf k})B^{-}({\mathbf k}')\rangle 
&=& (2\pi)^3 P_{\cal B}(k)\delta^{(3)}({\bf k} + {\bf k'}).
\end{eqnarray}
Note that the overall sign of the definition of these spectra depends on
the choice of the polarization vector. 

Using these power spectra, one can write the angle
dependence of the bispectrum of curvature perturbations as
\cite{shiraishi:2012}  
\begin{eqnarray}
B_\zeta ({{\bf k}_1}, {{\bf k}_2}, {{\bf k}_3}) 
&\propto& 
P_B(k_*) P_B(k_1) P_B(k_2) 
\left( \frac{1}{3} \mu^2_{12} + \mu^2_{23} + \mu^2_{31} 
- \frac{2}{3} - \mu_{12} \mu_{23} \mu_{31} \right) 
\nonumber \\  
&-&
P_B(k_*) P_{\cal B}(k_1) P_{\cal B}(k_2) 
\left( \mu_{23} \mu_{31} - \frac{1}{3} \mu_{12} \right) 
\nonumber \\  
&+& 
(1 \to 3, ~ 2 \to 1, ~ 3 \to 2) + (1 \to 2, ~ 2 \to 3, ~ 3 \to 1)
 ~, \label{eq:bis_curv_pmf} 
\end{eqnarray}
where $\mu_{ab}\equiv \hat{\bf k}_a\cdot \hat{\bf k}_b$ 
and $k_*$ denotes some pivot wavenumber.\footnote{Note that parity-odd
terms, which are proportional to $P_B^2 P_{\cal B}$ and $P_{\cal B}^3$,
do not appear in eq.~(\ref{eq:bis_curv_pmf}), as $\zeta$
is a scalar. On the other 
hand, the bispectrum involving vector or tensor perturbations may
contain parity-odd terms, which yield the CMB temperature
auto-bispectrum with $\ell_1 + \ell_2 + \ell_3 = {\rm odd}$
\cite{Shiraishi:2011st, shiraishi:2012}.} 

Eq.~(\ref{eq:bis_curv_pmf}) clearly shows that helical and non-helical
magnetic fields generate $L=1$ and $L=2$ angular dependence in the
bispectrum of curvature perturbations. If the magnetic field was
generated at a GUT scale, i.e., $\tau_\nu / \tau_B \approx 10^{17}$, with
nearly scale-invariant spectra of $P_B$ and $P_{\cal B}$, the Legendre
coefficients in eq.~(\ref{eq:bis_curv_general}) are related to the
amplitudes of non-helical and helical 
magnetic fields smoothed on $1~{\rm Mpc}$ as
\cite{shiraishi:2012} 
\begin{eqnarray}
c_0 \approx - 2 \times 10^{-4} 
\left(\frac{B_{1 \rm Mpc}}{\rm nG}\right)^6, \ \ 
 c_1 \approx - 0.9 
\left(\frac{B_{1 \rm Mpc}}{\rm nG}\right)^2
\left(\frac{{\cal B}_{1 \rm Mpc}}{\rm nG}\right)^4, \ \ 
 c_2 \approx 14 c_0.
\end{eqnarray} 
Therefore, if inflation creates $B_{1 \rm Mpc} \sim 3 {\rm nG}$ and
${\cal B}_{1 \rm Mpc} \sim 1 {\rm nG}$, which are consistent with the
current observational limits, we may have negative and non-vanishing
$c_1$ and $c_2$; namely, $c_1 \sim - 8$ and $c_2 \sim -2$. 
 
\subsection{$I^2(\phi)F^2$ model}
\label{subsec:vectors}

Vector fields with the standard Maxwell $- F^2/4$ kinetic term are not
produced by the expansion of the universe and, if generated by some
other source, they are rapidly diluted away. This poses a challenge to
models of primordial magnetogenesis and of vector fields during
inflation. 
Vector fields during inflation can result in broken statistical isotropy
of the primordial perturbations, which will be probed 
by the forthcoming {\sl Planck} data \cite{Pullen:2007tu,Ma:2011ii}. 

Vector fields with a kinetic term given by
\begin{equation}
{\cal L} = - \frac{I^2 \left( \phi \right)}{4} \, F^2 \;,
\label{ratra}
\end{equation}
can  instead  be  produced during inflation if $I \left( t \right)$ has
an appropriate time dependence \cite{Ratra:1991bn}. It is convenient to
define the ``electric'' and ``magnetic''
components
\begin{equation}
E_i = - \frac{I}{a^2} A_i'\;\;,\;\; B_i = \frac{I}{a^2} \epsilon_{ijk}
 \partial_j A_k\;,
\label{EB-ratra}
\end{equation}
where the primes denote derivatives with respect to the conformal time, and $a$ is the scale factor of the universe. In terms of these components, 
 the physical energy density in the vector field assumes the
 conventional expression,  
$\rho_A = \frac{\vert {\bf E} \vert^2+\vert {\bf B} \vert^2}{2}$. 

If $I \propto a^n$, with $n=+2$ or $n=-3$, the magnetic modes are
generated during inflation with a scale invariant  and frozen spectrum
outside the horizon. Rather than assuming that $I$ is an external
function, one can obtain the required time dependence by assuming that
$I$ is a function of the inflaton $\phi$, with a functional form related
to the inflaton potential by \cite{Ratra:1991bn,Martin:2007ue} 
 \begin{equation}
I = I_0 \, {\rm exp } \left[ - \int \frac{n \, d \phi}{\sqrt{2 \epsilon \left( \phi \right)} M_p} \right] \;\;\;\Rightarrow\;\;\;
\langle I \rangle \propto a^n,
\label{I-V}
\end{equation}
where $\epsilon$ is the usual slow-roll parameter, $\epsilon \equiv
\frac{M_p^2}{2} \, \left( \frac{1}{V} \frac{d V}{d \phi}\right)^2$, with
$M_p \equiv 1/\sqrt{8 \pi G}$ denoting the reduced Planck mass. 

Some recent work studied whether this coupling can result in visible cross-correlations between  primordial perturbations and 
 large-scale magnetic fields
 \cite{Caldwell:2011ra,Motta:2012rn,Jain:2012ga,Jain:2012vm,shiraishi/saga/yokoyama:2012,kunze:2013,kunze:prep}. This is
 not trivial to realize, as the $n=-3$ choice results in  too large an
 energy density in the electric modes \cite{Demozzi:2009fu,Kanno:2009ei}, while
 $n=+2$ leads to too large an electromagnetic coupling constant during inflation \cite{Demozzi:2009fu,Barnaby:2012tk}.\footnote{These problems persist also for a general evolution of  $I$ beyond the $a^n$ scaling \cite{Fujita:2012rb}.}
 
 For these reasons, we prefer not to identify the vector field as the
 electromagnetic field, and we discuss this model only as a mechanism
 for producing 
 non-Gaussianity.\footnote{We continue to adopt  the ``electromagnetic''
 decomposition (\ref{EB-ratra}) for notational convenience.} Indeed the
 vector modes are coupled to the inflaton field by the same $I^2 \left(
 \phi \right) F^2$ term that generates them, and  they source inflaton
 perturbations through this coupling. These perturbations add up
 incoherently to the inflaton vacuum modes, and are highly non-Gaussian.

 Ref.~\cite{Seery:2008ms} computed the resulting bispectrum, $\langle
 \zeta_{{\bf k}_1} \zeta_{{\bf k}_2} \zeta_{{\bf k}_3} \rangle$, in the
 equilateral configuration, i.e., $k_1=k_2=k_3$. The full
 bispectrum was computed in
 refs.~\cite{Barnaby:2012tk,Bartolo:2012sd}. The computations of
 refs.~\cite{Barnaby:2012tk,Bartolo:2012sd} are restricted to   $n=2$
 and $n=-2$ which produce, respectively, scale invariant ``magnetic''
 and ``electric'' perturbations. The model enjoys a symmetry $f
 \leftrightarrow \frac{1}{f}$, or $n \leftrightarrow - n$, under which
 $\vert E \vert \leftrightarrow \vert B \vert$. Both $\pm n$ result in
 the same equation for $\zeta$. For brevity of exposition, we only refer
 to the $n=-2$ case in the reminder of this subsection. In this case one
 obtains, at the leading order in slow-roll,
\begin{equation}
E_k \simeq \frac{3 H^2}{\sqrt{2} k^{3/2}} \;\;,\;\; B_k \simeq \frac{H}{\sqrt{2} k^{1/2} a} \;\;,\;\; k \ll a H
\label{Ek-Bk}
\end{equation}
for the mode functions of each of the two polarizations of  the
``electric'' and the ``magnetic'' fields in the super-horizon regime
\cite{Bartolo:2012sd}. We note that the power in the ``electric'' field
is frozen outside the horizon and scale invariant, whereas the power in the ``magnetic'' field  decreases to negligible values. 

Let us assume that, at the beginning of inflation, say at the time
$t=0$,  the ``electric'' field has a classical  homogeneous value, ${\bf
E}^0$, all across the universe with negligible perturbations.  For
$n=-2$, the classical equations of motion for the vector field are
solved by a constant, ${\bf E} = {\bf E}^0$. This quantity is, however,
not the classical ``electric'' field that would be measured by a local
observer at $t > 0$, which we denote by ${\bf E}_{\rm cl}$. In fact, the
modes given by eq.~(\ref{Ek-Bk}) become classical after they leave the
horizon (we 
denote them as infra-red (IR) modes), and they add up with ${\bf E}^0$ to
give  ${\bf E}_{\rm cl}$. A given IR mode of wavelength $\lambda$
averages to zero on regions of size $L \gg \lambda$, but it is constant
in each region  of the size $L \ll \lambda$, and it adds up stochastically with  ${\bf E}^0$ and with all the other modes with $\lambda \gg L$ generated during inflation to determine the value  ${\bf E}_{\rm cl}$ in that region.
An observer at time $t>0$ during inflation can only experience the value
of ${\bf E}_{\rm cl}$ in its local Hubble patch. The average measured by
this observer is drawn from a Gaussian distribution with the mean ${\bf
E}^0$ and the variance given by
\begin{equation}
\langle {\bf E}\cdot{\bf E} \rangle = \frac{2 \times 4 \pi}{\left( 2 \pi \right)^3} \int_{H a \left( t=0 \right)}^{H a \left( t \right)} \frac{d k}{k} k^3 E_k^2 = \frac{9 H^4}{2 \pi^2} \, N,
\label{variance}
\end{equation}
where $N$ is the number of e-folds from the start of inflation to the
time $t$.  The lower limit in the integral  corresponds to the modes that
left the horizon at the start of inflation (larger modes are not
generated), while the upper limit corresponds to the modes that left the
horizon at the time $t$ (larger-momentum modes are still in the quantum
regime and 
do not contribute to the classical average, 
${\bf E}_{\rm cl}$, in the Hubble patch of length $ \frac{1}{a\left( t
\right) H} $). 

The situation is completely identical to what happens to the so-called
``stochastic inflation  \cite{Starobinsky:1986fx},'' in which the
variance of a massless
scalar field, $\chi$, is determined by the stochastic
addition of the IR modes, and grows as $\langle \chi^2 \rangle \propto
H^2 N$ during inflation. It is well established in that context that
this variance contributes to the theoretical expectation value of the
scalar field measure by local observers. This is customary used, for
instance, in the Affleck-Dine model of baryogenesis
\cite{Affleck:1984fy} or in the curvaton field \cite{Lyth:2001nq}. The
fact that the vector field has spin $1$ does not make any difference for
these considerations, which simply follow from eq.~(\ref{Ek-Bk}). 

Let us consider a mode, $\zeta_k$, of a given comoving momentum, $k$. This mode leaves the horizon  $N_k $ e-folds before the end of inflation. 
We are interested in the modes that affect the CMB anisotropies. Such
modes leave the horizon $N_k \simeq N_{\rm CMB} \simeq 60$ e-folds
before the end of inflation. The Hubble patch that they exit is the one
that eventually becomes our Hubble patch. When the mode $\zeta_k$ leaves
the horizon,  the classical average of the ``electric'' field, ${\bf
E}_{\rm cl}$, in this Hubble patch is drawn from a Gaussian distribution
with the mean ${\bf E}^0$ and the variance given by
eq.~(\ref{variance}), with $N=N_{\rm tot} - N_k$, where $N_{\rm tot}$ is
the total number of e-folds of inflation. 

In the presence of this mean, the kinetic term given in
eq.~(\ref{ratra}) results in the coupling of $ {\cal L}_{\rm
int} \simeq 4 a^4 {\bf E}_{\rm cl} \cdot \delta {\bf E} \, \zeta$. This
is the dominant operator for the part of $\zeta$ sourced by the vector
field \cite{Bartolo:2012sd}.  The power spectrum of $\zeta$ generated by
this mechanism is given by
\begin{equation}
P_\zeta(k) 
= 
 P_\zeta^{(0)} \left( k \right) \left[ 1 + g_* \left( k \right) \cos^2 \theta_{{\hat k},{\hat E}_{\rm cl}} \right] \;\;,\;\; g_* \left( k \right) \simeq - \frac{ 24 E_{\rm cl}^2 \, N_k^2}{\epsilon \, V \left( \phi \right)}\;\;,
\label{z2-phiF2}
\end{equation}
where $ P_\zeta^{(0)} \equiv \frac{2 \pi^2}{k^3} \frac{H^2}{8 \pi^2 \epsilon M_p^2}$ is the square amplitude  of the standard vacuum modes. The second term 
in eq.~(\ref{z2-phiF2}) is the contribution of the sourced part of
$\zeta$, which (as phenomenologically required) we have assumed to be
subdominant.\footnote{Non-detection of statistical anisotropy
in the {\sl 
WMAP} 9-year data after the correction of non-circular beam effects
\cite{bennett/etal:prep} would imply a conservative 
upper bound of $g_*\lesssim 0.1$.} 
It follows from the $N_k^2$ proportionality that this term
continues to grow in the super-horizon regime. 
This power spectrum was
previously obtained in
refs.~\cite{Dulaney:2010sq,Gumrukcuoglu:2010yc,Watanabe:2010fh} in the
context of the anisotropic inflationary model \cite{Watanabe:2009ct},
where however $E_{\rm cl}$ was identified with ${\bf E}^0$, missing the
IR contribution.

The $\delta E \, \zeta$ mixing results in the following bispectrum 
in the squeezed limit, $k_3 \ll k_1 \approx k_2$
\cite{Bartolo:2012sd}\footnote{See ref.~\cite{Bartolo:2012sd} for the
full expression; due to different conventions, the bispectrum of
ref.~\cite{Bartolo:2012sd} is the one given here divided by $\left( 2
\pi \right)^{3/2}$.} 
\begin{eqnarray}
B_\zeta ( k_1, k_2, k_3 ) 
& \simeq & 
 24 P_\zeta^{(0)} \left( k_1 \right)  P_\zeta^{(0)} \left( k_3 \right) 
\, \vert g_* \left( k_1 \right) \vert N_{k_3} 
\nonumber \\ 
&\times &
  \Bigg[  
 1 - \cos^2 \theta_{ {\hat k}_1 ,\, {\hat E}_{\rm cl} } - \cos^2 \theta_{ {\hat k}_3 ,\, {\hat E}_{\rm cl} } 
 +   \cos \theta_{ {\hat k}_1 ,\, {\hat E}_{\rm cl} } \cos \theta_{ {\hat k}_3 ,\, {\hat E}_{\rm cl} } \cos \theta_{ {\hat k}_1 ,\, {\hat k}_3 } 
\Bigg]. 
\label{z3-phiF2}
 \end{eqnarray}

The predicted power spectrum (eq.~\ref{z2-phiF2}) and bispectrum
(eq.~\ref{z3-phiF2}) break statistical isotropy, as ${\bf E}_{\rm cl}$ picks
out a preferred direction. However, the prediction for an isotropic
measurement is obtained by averaging eq.~(\ref{z3-phiF2}) over all
directions of $ {\hat E}_{\rm cl} $.\footnote{For studies of the
bispectrum without averaging over preferred directions, see
refs.~\cite{Shiraishi:2011ph, Bartolo:2011ee}.}
We find
\begin{equation}
B_\zeta ( k_1, k_2, k_3 ) 
\Big\vert_{\rm isotropic \; measurement} \simeq 
     8  P_\zeta^{(0)} \left( k_1 \right)  P_\zeta^{(0)} \left( k_3 \right)     \,
  \vert g_* \left( k_1 \right) \vert N_{k_3}
  \left( 1 + \mu_{13}^2\right),
\label{z3-phiF2-iso}
\end{equation}
where $\mu_{13}\equiv \hat{\bf k}_1\cdot\hat{\bf k}_3$.
From this we obtain
 the Legendre
coefficients in eq.~(\ref{eq:bis_curv_general}) as
\begin{equation}
c_0 = 32  \, \frac{\vert g_* \left( k_1 \right) \vert}{0.1} \,  \frac{N_{k_3}}{60}  \;\;\;,\;\;\;
c_2 = \frac{c_0}{2}.
\label{c0-c2-phiF2}
\end{equation}

Due to simplicity of the model, eq.~(\ref{c0-c2-phiF2}) is a 
 very predictive
result, relating the bispectrum coefficients to the amount of
statistical anisotropy of the power spectrum, i.e.,
$g_*$. This result holds for all models characterized by eq.~(\ref{ratra})
and scale invariant ``magnetic'' or ``electric'' modes, including many
 analyses of the magnetogenesis mechanism \cite{Ratra:1991bn} and of
 anisotropic inflation \cite{Watanabe:2009ct} (for which the departure
 from scale invariance is negligibly small; we note that in this work
 ${\bf E}^0$ evolves on an attractor solution, but this has no
 consequence for the accumulation of the IR modes). An analogous result
 will also hold for the model of ref.~\cite{Emami:2013bk} and for the
 mechanism of ref.~\cite{Dimopoulos:2009vu}, for which the scalar
 perturbations have been studied in ref.~\cite{Namba:2012gg}.

The smallness of $g_*$ limits the level of non-Gaussianity. However, a
larger bispectrum, for a given value of $g_*$, can be obtained if the
model is more complicated. For instance, one can arrange for a triplet
of U(1) vectors, and assume that they have classical vacuum expectation values
 which are orthogonal to one another and of equal magnitudes \cite{Funakoshi:2012ym}. In this case the power spectrum is statistically isotropic ($g_* =0$).
This requires to assume that the IR sum is subdominant, as there is no
reason to assume that the IR modes of the three vectors add up to
orthonormal values. A larger bispectrum can also be obtained if there
are additional fields and additional couplings, as in the waterfall
mechanism of ref.~\cite{Yokoyama:2008xw}, in which a vector field of the
kinetic term given in eq.~(\ref{ratra}) is also coupled to the field
that determines the end of hybrid inflation (see
ref.~\cite{Bartolo:2012sd} for more detailed discussion). 

\subsection{Solid inflation}
\label{subsec:solid}

Ref.~\cite{Endlich:2012pz} studied   a rather unusual model, in which
 inflation is driven by a system which has a field-theoretical
 description of a solid.   An equivalent version of the model was
 proposed by ref.~\cite{Gruzinov:2004ty}, under the name of ``elastic
 inflation.'' 

Each volume element of the solid is characterized by a comoving label,
$\phi^i$ (for instance, it can be the position of that element at the
initial time $t=0$). The functions, $\phi^i \left( t , {\bf x} \right)$,
specify which volume element is located at a given position, ${\bf x}$,
at a given time, $t$. A  solid at rest in comoving coordinates then obeys
\begin{equation}
\langle \phi^1 \rangle = x \;\;,\;\; \langle \phi^2 \rangle = y \;\;,\;\; \langle \phi^3 \rangle = z
\label{vevs}
\end{equation}
or, in short, $\langle \phi^i \rangle = x^i$. Even if the vacuum
expectation value of each
field is ${\bf x}$-dependent, a homogeneous and isotropic
Friedmann-Robertson-Walker solution can still be obtained by requiring
that the Lagrangian that controls the solid be invariant under rigid
translations,   $\phi^i \rightarrow \phi^i + a^i$, and SO(3) rotations,
$\phi^i \rightarrow O^i_j \phi^j$.         

At the lowest order in a derivative expansion, the translational
invariance is guaranteed by considering functions
\begin{equation}
B^{ij} = g^{\mu \nu} \partial_\mu \phi^i \partial_\nu \phi^j,
\label{Bij}
\end{equation}
and isotropy is  obtained by requiring that the Lagrangian is a function of SO(3) invariants built from $B^{ij}$. Only three  independent invariants
exist, and ref.~\cite{Endlich:2012pz} chose
\begin{equation}
S_{\rm solid} = \int d^4 x \sqrt{-g} F \left[ X , Y , Z \right] \;\;,\;\;
X \equiv \left[  B \right] \;\;\;,\;\; 
Y \equiv \frac{ \left[  B^2 \right] }{\left[ B \right]^2} \;\;,\;\;
Z \equiv \frac{ \left[  B^3 \right] }{\left[ B \right]^3} \;,
\end{equation}
where the square parenthesis denotes the trace of the corresponding
matrix, e.g.,   $\left[ B \right] \equiv \sum_i B^{ii}$.

The system  has the energy-momentum tensor, $T^\mu_{~\mu} = {\rm diag } \left( - \rho , p , p , p \right)$, with \cite{Endlich:2012pz}   
\begin{equation}
\rho = - F \;\;,\;\; p = F - \frac{2}{a^2} F_X \;,
\end{equation}
where the subscript denotes a partial derivative, and $F$ and $F_X$ are
   evaluated on the background solutions given by $X = \frac{3}{a^2 \left( t
   \right)}$, $Y = \frac{1}{3}$, and $Z = \frac{1}{9}$. These invariants
   are chosen in such a way that $X$ is the only one affected by the
   overall physical volume expansion; this immediately explains as to why only
   the derivative of $F$ with respect to $X$ enters into the expression
   for the pressure. 

   Inflation is possible only if $F$ is only mildly affected by the physical expansion, or, equivalently, only if $F_X$ is sufficiently small. Specifically, we require 
 \begin{equation}
 \epsilon \equiv - \frac{\dot{H}}{H} = \frac{X F_X}{F} \ll 1.
 \end{equation}
One also requires $F_{XX}$ to be small, so that $\eta \equiv \frac{\dot{\epsilon}}{\epsilon H} \ll 1$ \cite{Endlich:2012pz}.   

Let us now discuss  cosmological perturbations in this system. It is
convenient to work in a spatially flat gauge, where the dynamical 
scalar and vector perturbations are all encoded in the perturbations of the scalar fields:
\begin{equation}
\delta \phi^i = \pi^i \left( t ,\, {\bf x} \right) = \frac{\partial_i }{\sqrt{-\nabla^2}} \, \pi_L + \pi_T^i \;,
\end{equation}
where the vector components are transverse, $\partial_i \pi_T^i =
 0$. The scalar and vector modes are decoupled from each other at the
 linearized level.\footnote{As always in cosmology, the sectors of scalar
 and vector perturbations also include non-dynamical modes which are, in
 the   spatially flat gauge, encoded    in the $\delta g_{0\mu}$ metric
 components. These fields can be integrated out as explained in
 ref.~\cite{Endlich:2012pz}. This affects the action for the dynamical modes,
 $\pi_L$ and $\pi_T^i$, at long wavelengths.  Finally, there are also
 tensor perturbations - the gravity waves - encoded in the $\delta
 g_{ij}$ metric components, and which we do not discuss here.} At the
 lowest order in the slow roll parameters, and in the deep sub-horizon
 regime, the sound speeds of the scalar (longitudinal) modes, $c_L$, and
 vector (transverse) modes, $c_T$,
 are given, 
 respectively, by  \cite{Endlich:2012pz}   
\begin{equation}
c_L^2 \simeq \frac{1}{3} + \frac{8}{9} \frac{F_Y+F_Z}{X F_X} \;\;,\;\;
 c_T^2 \simeq \frac{3}{4} \left( 1 + c_L^2 \right) \;,
\end{equation}
so that the propagation is subluminal  and non-tachyonic ($0 < c_L^2<1$
and $0<c_T^2 < 1$) for $0 < F_Y + F_Z < \frac{3}{8} \epsilon \vert F
\vert$~\cite{Endlich:2012pz}.
Namely, the requirement of an  accelerated expansion forces $F_X$ to be
small, while subluminality also requires that the combination $F_Y +
F_Z$ be small. Finally, the theory involves derivative interactions of
the ``phonon'' fields, $\pi^i$, which necessarily become strong at some
scale $\Lambda$. A detailed study in ref.~\cite{Endlich:2012pz} gives an
estimate of $\Lambda \gg H$ (so that     
 the linearized theory is also valid in the sub-horizon regime, up to
 $\Lambda$),  provided that $\epsilon c_L^3 \gg \left( \frac{H}{M_p}
 \right)^{2/3}$, which can always be obtained for a sufficiently small $H$.
 
 In a conformally  flat gauge, the   gauge-invariant scalar perturbation, $\zeta$, evaluates to  $\zeta = - H \frac{\delta \rho}{\dot{\rho}}
  = \frac{1}{3} \partial \pi$. Its solution  exhibits two features that
  are peculiar in scalar-field inflation models, but that were
  nevertheless already seen in the models studied in the previous
  subsection \cite{Barnaby:2012tk}. The first one is the fact that
  $\zeta$ is not conserved on super-horizon scales, due to the
  anisotropic stress that does not vanish on super-horizon scales
  \cite{Endlich:2012pz}. Indeed, following ref.~\cite{Endlich:2012pz},
  we obtain 
\begin{equation}
\delta T_{ij,{\rm scalar}} =  a^2  M_p^2 \dot{H} \zeta  \left[ 2 \left(3   - 2 \epsilon + \eta \right) \delta_{ij} - \left( 3 + 3 c_L^2 - 2 \epsilon + \eta \right) \left( 3 {\hat k}_i \, {\hat k}_j - \delta_{ij} \right) \right].
\label{Tij}
\end{equation}

Let us recall that, also for the model described by eq.~(\ref{ratra}), 
the anisotropic component of the stress-energy tensor, $\propto E_{{\rm
cl},i} \delta E_j$, does not vanish outside the horizon, and thus the
anisotropic term in eq.~(\ref{z2-phiF2}) grows outside the horizon.

The second feature is that, analogously to the model described by
eq.~(\ref{ratra}), the bispectrum of solid inflation is largest in the
squeezed configurations, and  exhibits a nontrivial dependence on the
angle between the modes in the squeezed limit. The dominant
contribution to the bispectrum is given by the interactions of $\pi$
encoded in the scalar field Lagrangian, while the metric perturbations
provide a negligible contribution. The dominant interaction in a slow
roll expansion is given by \cite{Endlich:2012pz}
\begin{equation}
{\cal L} \supset M_p^2 a^3 H^2 \frac{F_Y}{F} \left[ \frac{7}{81} \left( \partial \pi \right)^2 - \frac{1}{9} \partial \pi \partial_j \pi^k \partial_k \pi^j 
- \frac{4}{9} \partial \pi \partial_j \pi^k \partial_j \pi^k + \frac{2}{3} \partial_j \pi^i \partial_j \pi^k \partial_k \pi^i \right].
\end{equation}
Also in this respect, the situation is analogous to the model discussed
in the previous subsection, where the dominant contribution to the
bispectrum is obtained from eq.~(\ref{ratra}), disregarding metric
perturbations. When written in terms of $\zeta$, this interaction
exhibits nontrivial dependence on the direction of the modes which does
not vanish in the squeezed limit, imprinting  the nontrivial
angular dependence in the bispectrum.  

At the leading order,  the Legendre
coefficients in eq.~(\ref{eq:bis_curv_general}) are given by
\cite{Endlich:2012pz} 
\begin{equation}
c_0 \simeq 0 \;\;,\;\; c_2 = {\cal O } \left( 1 \right) \frac{F_Y}{F} \frac{1}{\epsilon c_L^2}.
\end{equation}
Namely, in the squeezed limit, the dominant contribution is given by the
  quadrupole term, whereas the monopole term is negligible.  
  The dominant quadrupole term is essentially proportional to a free
  combination of parameters (we recall that avoiding superluminality and
  strong coupling at $p \la H$ imposes restrictions on the combination
  $F_Y + F_Z$ but not on $F_Y$ or $F_Z$ individually).

\section{Signatures in the cosmic microwave background}\label{sec:CMB_bis}

In this section, we shall derive the flat-sky (section~\ref{sec:flat}) and
full-sky (section~\ref{sec:full}) formulae for the bispectrum of CMB
temperature anisotropy from the bispectrum  of  curvature perturbations
given in eq.~(\ref{eq:bis_curv_general}). 
We then calculate, in section~\ref{sec:2DFisher},
 the error bars of $c_0$, $c_1$, and
$c_2$ expected for a cosmic-variance-limited experiment measuring
temperature anisotropy up to $\ell=2000$.

\subsection{Flat-sky formula}
\label{sec:flat}
While the full-sky formula is eventually needed for the analysis of
full-sky temperature maps, let us derive first the flat-sky formula, as
the flat-sky formula is usually simpler and more intuitively understandable.

Under the flat-sky approximation, which is valid only on sufficiently
small angular scales, $\ell \gg 1$, 
CMB fluctuations on the sky are
expanded using the two-dimensional Fourier transform, instead of the
spherical harmonics. The Fourier coefficients of CMB anisotropy,
$a({\boldsymbol \ell})$ are related to the three-dimensional 
coefficients of the curvature perturbation, $\zeta({\bf k})$, as \cite{shiraishi/etal:2010a} 
\begin{eqnarray}
a({\boldsymbol \ell}) &=& \int_0^{\tau_0} d\tau \int_{-\infty}^{\infty}
 \frac{dk_z}{2 \pi} \zeta \left({\bf k}^{\parallel} = -\frac{\boldsymbol
			   \ell}{D}, k_z\right) 
 S_I\left(k = \sqrt{k_z^2 + (\ell/D)^2},\tau\right) \frac{1}{D^2}e^{- i k_z D} ~,
 \label{eq:flat_alst} 
\end{eqnarray}
where ${\bf k}^{\parallel} \equiv (k_x, k_y)$, $D \equiv \tau_0 - \tau$
is the conformal distance out to a given epoch $\tau$, $\tau_0$ is the
present-day conformal time, and $S_I$ is the so-called source
function.\footnote{The source function is related to the radiation transfer
function defined in eq.~(\ref{eq:gT}) as ${\cal T}_\ell(k) =
\int_0^{\tau_0} d\tau~S_I(k,\tau) j_\ell(kD)$.} 
This relation simply tells us that $a({\boldsymbol\ell})$ measures the
			   $\zeta$ modes that are perpendicular to the
			   line-of-sight direction (i.e., the modes on the sky),
			   and the line-of-sight modes are washed out by
			   integration.  

It is straightforward to compute the bispectrum of $a({\boldsymbol\ell})$
following, e.g., ref.~\cite{shiraishi/etal:2010a}:
\begin{eqnarray}
\left\langle a({\boldsymbol \ell_1})a({\boldsymbol \ell_2})a({\boldsymbol \ell_3}) \right\rangle
 = (2\pi)^2 
\delta^{(2)}
\left({\boldsymbol \ell}_1 + {\boldsymbol \ell}_2 + {\boldsymbol \ell}_3\right) 
\sum_L c_L b^{L}(\ell_1, \ell_2, \ell_3) ~,   
\end{eqnarray}
where
\begin{eqnarray}
b^{L}(\ell_1, \ell_2, \ell_3) &\equiv& 
\int_{-\infty}^\infty r^2 dr \left[ \prod_{n=1}^3 \int_0^{\tau_0} d\tau_n \int_{\ell_n / D_n}^{\infty}
 \frac{d k_n}{2 \pi} 
{\cal G}(\ell_n, k_n, \tau_n, r) \right] \nonumber \\ 
&&\times 
\left[
\sum_{n=0}^L 
( \hat{\boldsymbol \ell}_1 \cdot \hat{\boldsymbol \ell}_2 )^n 
{\cal F}_L^{(n)} \right]
P_\zeta(k_1) P_\zeta(k_2) 
 + \mbox{(2 perm)} ~,  
\label{eq:originalflat}
\end{eqnarray}
with 
\begin{equation}
{\cal G}(\ell, k, \tau, r) 
\equiv \left[ 1 - \left( \frac{\ell}{k D} \right)^2 \right]^{-1/2}
S_I(k, \tau) \frac{2}{D^2} 
\cos \left[\sqrt{1 - \left( \frac{\ell}{k D} \right)^2} 
k (r - D) \right]  ~.
\end{equation}
The other kernel functions, ${\cal F}_L^{(n)}$, for $L\le 2$ are given by
${\cal F}_0^{(0)} = 1$,  
${\cal F}_1^{(1)}= \prod_{n=1}^2 \frac{\ell_n}{k_nD_n}$, 
${\cal F}_2^{(2)}
= \frac{3}{2}  \prod_{n=1}^2 \left( \frac{\ell_n}{k_n D_n} \right)^2$, and
\begin{eqnarray}
{\cal F}_1^{(0)}
&=& - \prod_{n=1}^2 \sqrt{1 - \left( \frac{\ell_n}{k_n D_n} \right)^2}
\tan \left[\sqrt{1 - \left( \frac{\ell_n}{k_n D_n} \right)^2}
 k_n (r - D_n)\right]  
~, \\ 
{\cal F}_2^{(0)}
&=& \frac{3}{2} \prod_{n=1}^2 
\left[ 1 - \left( \frac{\ell_n}{k_n D_n} \right)^2 \right]  - \frac{1}{2} 
~, \\  
{\cal F}_2^{(1)}
&=& - 3  \prod_{n=1}^2 \frac{\ell_n}{k_n D_n} 
\sqrt{1 - \left( \frac{\ell_n}{k_n D_n} \right)^2}
\tan \left[\sqrt{1 - \left( \frac{\ell_n}{k_n D_n} \right)^2}
 k_n (r - D_n)\right]  
~. 
\end{eqnarray}

While these formulae are still complicated, one can read off the leading
behaviours of these expressions in the small-scale
limit, in which the dominant contributions in the $k$ integration come from the
modes with $k\approx \ell/D$. In this limit the kernel functions become
${\cal F}_1^{(0)}\to 0$, ${\cal F}_2^{(1)} \to 0$, ${\cal F}_1^{(1)} \to
1$, ${\cal F}_2^{(0)} \to -1/2$, and ${\cal F}_2^{(2)} \to 3/2$. We thus
find
\begin{eqnarray}
b^{L}(\ell_1, \ell_2, \ell_3) &\to& 
\int_{-\infty}^\infty r^2 dr \left[ \prod_{n=1}^3 \int_0^{\tau_0} d\tau_n \int_{\ell_n / D_n}^{\infty}
 \frac{d k_n}{2 \pi} 
{\cal G}(\ell_n, k_n, \tau_n, r) \right] \nonumber \\ 
&&\times 
P_L( \hat{\boldsymbol \ell}_1 \cdot \hat{\boldsymbol \ell}_2 )
P_\zeta(k_1) P_\zeta(k_2) 
 + \mbox{(2 perm)}.
\label{eq:flatskybispectrum}
\end{eqnarray}
This result shows that the CMB bispectrum is proportional to
$P_L(\hat{{\boldsymbol \ell}_1} \cdot 
\hat{{\boldsymbol \ell}_2})$ (and its permutations),
which is expected from  
$P_L(\hat{{\bf k}}_1 \cdot \hat{{\bf
k}}_2)$ (and its
permutations) in the
three-dimensional bispectrum of the primordial curvature
perturbation. 

While the approximation of $\ell\approx kD$ gives a simple and
transparent result given in eq.~(\ref{eq:flatskybispectrum}), it is less
precise than the original form given by eq.~(\ref{eq:originalflat}). In
appendix~\ref{app:flat}, we discuss the precision of
eqs.~(\ref{eq:originalflat}) and (\ref{eq:flatskybispectrum}) with
respect to the full-sky result given in the next subsection.

\subsection{Full-sky formula}
\label{sec:full}
Encouraged by the flat-sky results, we now move onto the full-sky
case. The CMB temperature anisotropy on the celestial sphere is expanded
by the spherical harmonic function as $\delta T(\hat{\bf n})/T =
\sum_{\ell, m} a_{\ell m} Y_{\ell m}(\hat{\bf n})$, where $\hat{\bf n}$
is a three-dimensional unit vector pointing toward a given direction on
the sky. The spherical harmonics coefficients, $a_{\ell m}$, are related
to the primordial curvature perturbation as 
\begin{eqnarray}
\nonumber
a_{\ell m} &=& 4\pi (-i)^{\ell} \int \frac{d^3{\bf k}}{(2\pi)^3} {\cal
T}_{\ell}(k) \zeta_{\mathbf k}Y_{\ell m}^*(\hat{\mathbf k})\\
&=& 4\pi (-i)^{\ell} \int \frac{k^2 dk}{(2\pi)^3} {\cal T}_{\ell}(k) 
\zeta_{\ell m}(k),
\label{eq:gT}
\end{eqnarray}
where ${\cal T}_\ell(k)$ is the so-called radiation transfer function,
and we have defined the curvature perturbation expanded in spherical
harmonics as $\zeta_{\ell m}(k) \equiv \int d^2 \hat{\bf k} ~\zeta_{\bf k} Y_{\ell m}^* (\hat{\bf k})$.
Then, the bispectrum of $a_{\ell m}$ can be straightforwardly calculated as 
\begin{eqnarray}
\left\langle \prod_{n=1}^3 a_{\ell_n m_n} \right\rangle
= 
\left[ \prod_{n=1}^3 4\pi (-i)^{\ell_n} \int \frac{k^2_n dk_n}{(2\pi)^3} {\cal T}_{\ell_n}(k_n) \right] 
\left\langle \prod_{n=1}^3 \zeta_{\ell_n m_n}(k_n) \right\rangle
~, \label{eq:CMB_bis_form}
\end{eqnarray} 
with the bispectrum of $\zeta_{\ell m}$ related to $B_\zeta(k_1,k_2,k_3)$ as
\begin{eqnarray}
\left\langle \prod_{n=1}^3 \zeta_{\ell_n m_n}(k_n) \right\rangle 
= \left[ \prod_{n=1}^3 \int d^2\hat{\bf k}_n
Y_{\ell_n m_n}^* (\hat{\bf k}_n) \right] (2\pi)^3 
\delta^{(3)} \left( {{\bf k}_1} + {{\bf k}_2} + {{\bf k}_3} \right)B_{\zeta}(k_1, k_2, k_3). \label{eq:angle_bis_curv}
\end{eqnarray}
Using the bispectrum of $\zeta$ given in
eq.~(\ref{eq:bis_curv_general}), $B_\zeta(k_1,k_2,k_3)$ can be expanded as
\begin{equation}
B_{\zeta}(k_1, k_2,k_3) 
= P_\zeta(k_1)P_\zeta(k_2) \sum_L c_L
\frac{4\pi}{2L+1} \sum_M Y_{L M}^* (\hat{{\bf k}}_1) Y_{L M} (\hat{{\bf k}}_2) 
+ \mbox{(2 perm)}.
\end{equation}
Using the definition of the delta function, $\delta^{(3)}({\mathbf
k})=(2\pi)^{-3}\int d^3{\mathbf x}~e^{i\mathbf{k}\cdot\mathbf{x}}$, we
also expand the delta function as
\begin{eqnarray}
\delta^{(3)}\left( {{\bf k}_1} + {{\bf k}_2} +{{\bf k}_3}  \right) 
&=& 8 \int_0^\infty r^2 dr 
\left[ \prod_{n=1}^3 \sum_{L_n M_n} 
 (-1)^{\frac{L_n}{2}} j_{L_n}(k_n r) 
Y_{L_n M_n}^*(\hat{\bf k}_n) \right] \nonumber \\ 
&&\times 
\left(
  \begin{array}{ccc}
  L_1 & L_2 & L_3 \\
  M_1 & M_2 & M_3 
  \end{array}
 \right)I_{L_1 L_2 L_3}~, 
\end{eqnarray}
where the $2 \times 3$ matrix denotes the Wigner-3$j$ symbol, and the $I$
symbol is defined by
\begin{eqnarray}
I_{l_1 l_2 l_3}
\equiv \sqrt{\frac{(2 l_1 + 1)(2 l_2 + 1)(2 l_3 + 1)}{4 \pi}}
\left(
  \begin{array}{ccc}
  l_1 & l_2 & l_3 \\
  0 & 0 & 0
  \end{array}
 \right)~.
\end{eqnarray}

Now, performing the integrals of the spherical harmonics over $\hat{{\bf
k}}_1, \hat{{\bf k}}_2$, and $\hat{{\bf k}}_3$, and performing the summations over $M_1$, $M_2$, $M_3$, and $M$ as described in ref.~\cite{shiraishi/etal:2011a}, we obtain
\begin{eqnarray}
\left\langle \prod_{n=1}^3 \zeta_{\ell_n m_n}(k_n) \right\rangle
&= (2\pi)^3 
 B_{\zeta, \ell_1 \ell_2 \ell_3}(k_1, k_2, k_3) 
\left(
  \begin{array}{ccc}
  \ell_1 & \ell_2 & \ell_3 \\
  m_1 & m_2 & m_3 
  \end{array}
 \right)
~,  
\label{eq:angle_bis_curv_wigner}
\end{eqnarray}
where
\begin{eqnarray}
\nonumber
B_{\zeta, \ell_1 \ell_2 \ell_3}(k_1, k_2, k_3) 
&=& 8 \int_0^\infty r^2 dr
\left[ \prod_{n=1}^3 \sum_{L_n} 
 (-1)^{\frac{L_n}{2}} j_{L_n}(k_n r)  \right] 
I_{L_1 L_2 L_3} \\ 
\nonumber
&&\quad \times \sum_L c_L 
\frac{4\pi}{2L+1}  
I_{\ell_1 L_1 L}
 I_{\ell_2 L_2 L}
(-1)^{\ell_2 + L_1} \delta_{L_3, \ell_3}  \\ 
&&\quad \times 
\left\{
  \begin{array}{ccc}
  \ell_1 & \ell_2 & \ell_3 \\
  L_2 & L_1 & L 
  \end{array}
 \right\}
   P_\zeta(k_1) P_\zeta(k_2)  
+ \mbox{(2 perm)}
~. \label{eq:angle_bis_curv_comp}
\end{eqnarray}
Here, the $2 \times 3$ matrix enclosed by curly brackets denotes the
Wigner-6$j$ symbol. As the primordial bispectrum given by
eq.~(\ref{eq:bis_curv_general}) is rotationally invariant, the
bispectrum expanded in spherical harmonics must also be rotationally
invariant. This means that the dependence of the bispectrum on $m_1,
m_2$ and $m_3$ must be given by the Wigner-$3j$ symbol, as shown in
eq.~(\ref{eq:angle_bis_curv_wigner}). This property ensures 
rotational invariance of the CMB bispectrum. 

Substituting eqs.~(\ref{eq:angle_bis_curv_wigner}) and
(\ref{eq:angle_bis_curv_comp}) into eq.~(\ref{eq:CMB_bis_form}), we
finally obtain the full-sky formula for the CMB bispectrum: 
\begin{eqnarray}
\left\langle \prod_{n=1}^3 a_{\ell_n m_n} \right\rangle 
&=& \left(
  \begin{array}{ccc}
  \ell_1 & \ell_2 & \ell_3 \\
  m_1 & m_2 & m_3 
  \end{array}
 \right) 
 B_{\ell_1 \ell_2 \ell_3} 
= 
\left(
  \begin{array}{ccc}
  \ell_1 & \ell_2 & \ell_3 \\
  m_1 & m_2 & m_3 
  \end{array}
 \right) 
 \sum_{L} c_L B_{\ell_1 \ell_2 \ell_3}^{L} 
~, 
\end{eqnarray}
where
\begin{eqnarray}
 B_{\ell_1 \ell_2 \ell_3}^{L} &\equiv&  
\int_0^\infty r^2 dr 
\left[ \prod_{n=1}^3 \sum_{L_n} 
 (-1)^{\frac{\ell_n + L_n}{2}}  \right]   I_{L_1 L_2 L_3}
\beta_{\ell_1 L_1}(r) \beta_{\ell_2 L_2}(r) \alpha_{\ell_3}(r)  
 \nonumber \\ 
&&\times  
\frac{4\pi}{2L+1}  
I_{\ell_1 L_1 L}
 I_{\ell_2 L_2 L}
(-1)^{\ell_2 + L_1} \delta_{L_3, \ell_3}
\left\{
  \begin{array}{ccc}
  \ell_1 & \ell_2 & \ell_3 \\
  L_2 & L_1 & L 
  \end{array}
 \right\}
+ \mbox{(2 perm)}~,
\label{eq:CMB_bis_comp}
\end{eqnarray}
and 
\begin{eqnarray}
\label{eq:alpha}
\alpha_{\ell}(r) &\equiv& \frac{2}{\pi} 
\int_0^\infty k^2 dk~ {\cal T}_{\ell}(k) j_\ell(k r)  ~, \\ 
\label{eq:beta}
\beta_{\ell L}(r) &\equiv& \frac{2}{\pi} 
\int_0^\infty k^2 dk~ P_{\zeta}(k) {\cal T}_{\ell}(k) j_L(k r)  ~.
\end{eqnarray}
Owing to the selection rules of the Wigner symbols, $\ell_1$, $\ell_2$
and $\ell_3$ are constrained by parity invariance and the triangular condition:
\begin{eqnarray}
\ell_1 + \ell_2 + \ell_3 = {\rm even}~, \ \ |\ell_1 - \ell_2| \leq \ell_3 \leq \ell_1 + \ell_2 ~.
\end{eqnarray}
The former constraint is a consequence of the bispectrum of curvature
perturbations given by eq.~(\ref{eq:bis_curv_general}) being parity-even.
The summation ranges of $L_1$ and $L_2$ are also restricted to  
\begin{eqnarray}
L_n = |\ell_n - L|, |\ell_n - L| + 2, \cdots, \ell_n + L - 2,  \ell_n + L ~.
\end{eqnarray}

\begin{figure}[t]
\centering{
 \includegraphics[width =1\textwidth]{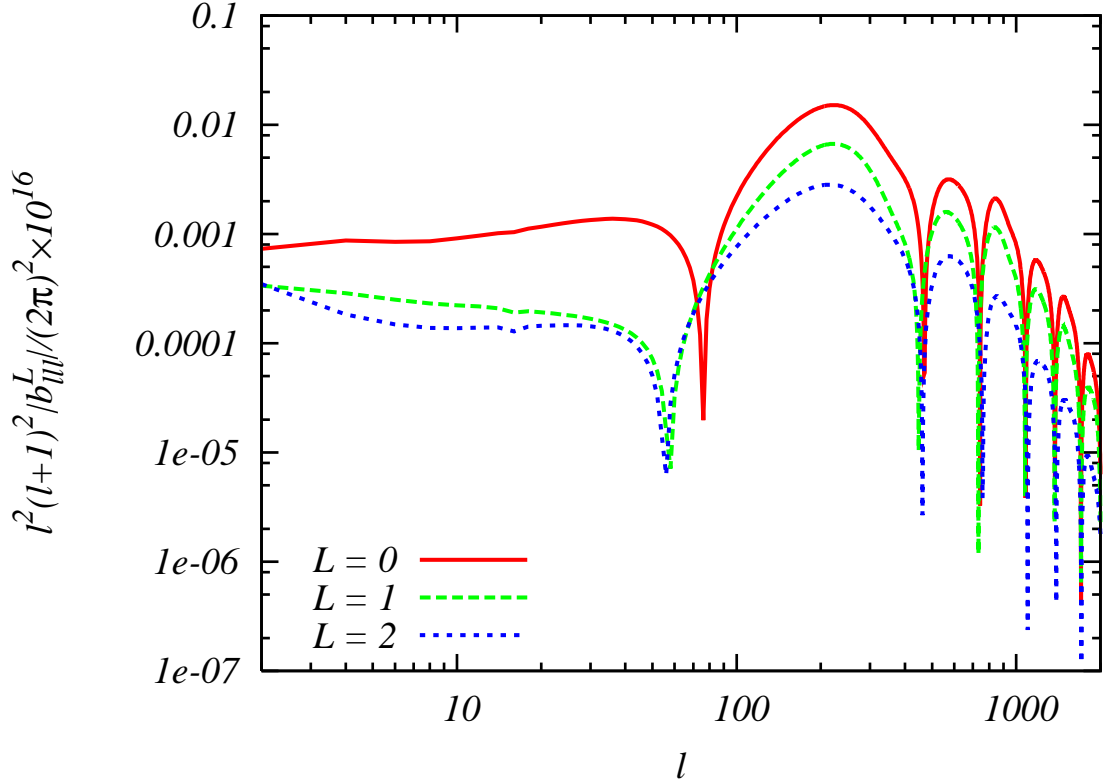}
}
  \caption{Absolute values of the {\it equilateral} CMB temperature
 reduced bispectra, $|b_{\ell\ell\ell}^L|$, for $L=0$
 (solid), $L=1$ (long-dashed), and $L=2$
 (short-dashed).}
  \label{fig:SSS_equ}
\end{figure} 
\begin{figure}[t]
\centering{
 \includegraphics[width =1\textwidth]{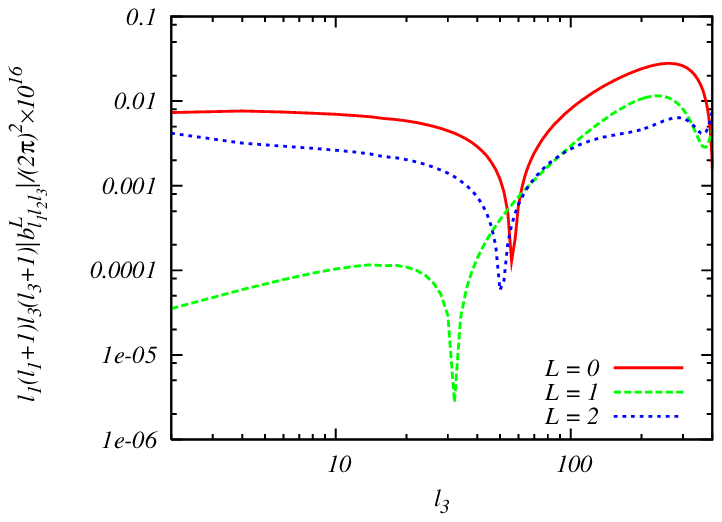}
}
  \caption{Same as figure~\ref{fig:SSS_squ}, but for the {\it squeezed}
 triangles, $|b_{\ell_1\ell_2\ell_3}^L|$, with $\ell_1 = \ell_2 = 200$, as
 a function of $\ell_3$.}
  \label{fig:SSS_squ}
\end{figure} 

In the full-sky formula given by eq.~(\ref{eq:CMB_bis_comp}), the angle
dependence for $L > 0$ induces a coupling among $\ell_1$,
$\ell_2$ and $\ell_3$ via the Wigner-$6j$ symbol. As a result,
eq.~(\ref{eq:CMB_bis_comp}) is not separable (or at least not
obviously separable) with respect to $\ell$'s
unlike the usual local-form CMB bispectrum without angle dependence,
i.e., $L=0$.

Figures~\ref{fig:SSS_equ} and \ref{fig:SSS_squ} show the absolute values
of the full-sky 
reduced bispectra, $b_{\ell_1 \ell_2 \ell_3}^{L}  
\equiv B_{\ell_1 \ell_2 \ell_3}^{L} 
\left( I_{\ell_1 \ell_2 \ell_3}\right)^{-1}$, 
for $L=0$, 1, and 2. Note that the full-sky reduced bispectrum reduces to the
flat-sky bispectrum, $b(\ell_1,\ell_2,\ell_3)$, that we discussed in the
previous subsection, in the small-sky limit \cite{Komatsu:2001rj}.

Figure~\ref{fig:SSS_equ} shows the equilateral triangles with
$\ell\equiv \ell_1 = \ell_2 = \ell_3$, while figure~\ref{fig:SSS_squ} shows
triangles with 
$\ell_1 = \ell_2 = 200$, which become squeezed triangles for $\ell_3\ll 200$.
We find that the amplitudes of the equilateral triangles monotonically
decrease as $L$ increases. We can understand this by using the flat-sky
formula given in eq.~(\ref{eq:flatskybispectrum}): the Legendre
polynomials give the ratio of $L=0$, 1, and 2 terms as
\begin{equation}
b^{L=0}(\ell,\ell,\ell):b^{L=1}(\ell,\ell,\ell):b^{L=2}(\ell,\ell,\ell)=1:-\frac12:-\frac18,
\end{equation}
for $\hat{\boldsymbol\ell}_i\cdot \hat{\boldsymbol\ell}_j=-\frac12$ ($i\neq j$).

Figure~\ref{fig:SSS_squ} shows the squeezed triangles with $\ell_3 \ll
\ell_1 = \ell_2=200$. In the squeezed limit, the CMB bispectrum of $L = 1$ is
highly suppressed compared with those of $L = 0$ and 2. This is simply due
to symmetry: the $L=1$ term vanishes in the exact squeezed limit.
Again, the flat-sky formula given in eq.~(\ref{eq:flatskybispectrum})
gives the ratio  of $L=0$, 1, and 2 terms as
\begin{equation}
b^{L=0}(\ell_1,\ell_1,\ell_3):b^{L=1}(\ell_1,\ell_1,\ell_3):b^{L=2}(\ell_1,\ell_1,\ell_3)=1:0:-\frac12,
\end{equation}
for $\hat{\boldsymbol\ell}_1\cdot \hat{\boldsymbol\ell_3}=0=\hat{\boldsymbol\ell}_2\cdot \hat{\boldsymbol\ell_3}$.

As these calculations are quite involved, we provide the simplified
analytical formula in the Sachs--Wolfe limit in
appendix~\ref{appen:SW}. This test validates our numerical
results shown in figures~\ref{fig:SSS_equ} and \ref{fig:SSS_squ}.

\subsection{Expected uncertainties on $c_1$ and $c_2$} \label{sec:2DFisher}
\begin{figure}[t]
  \begin{center}
    \includegraphics[width =1\textwidth]{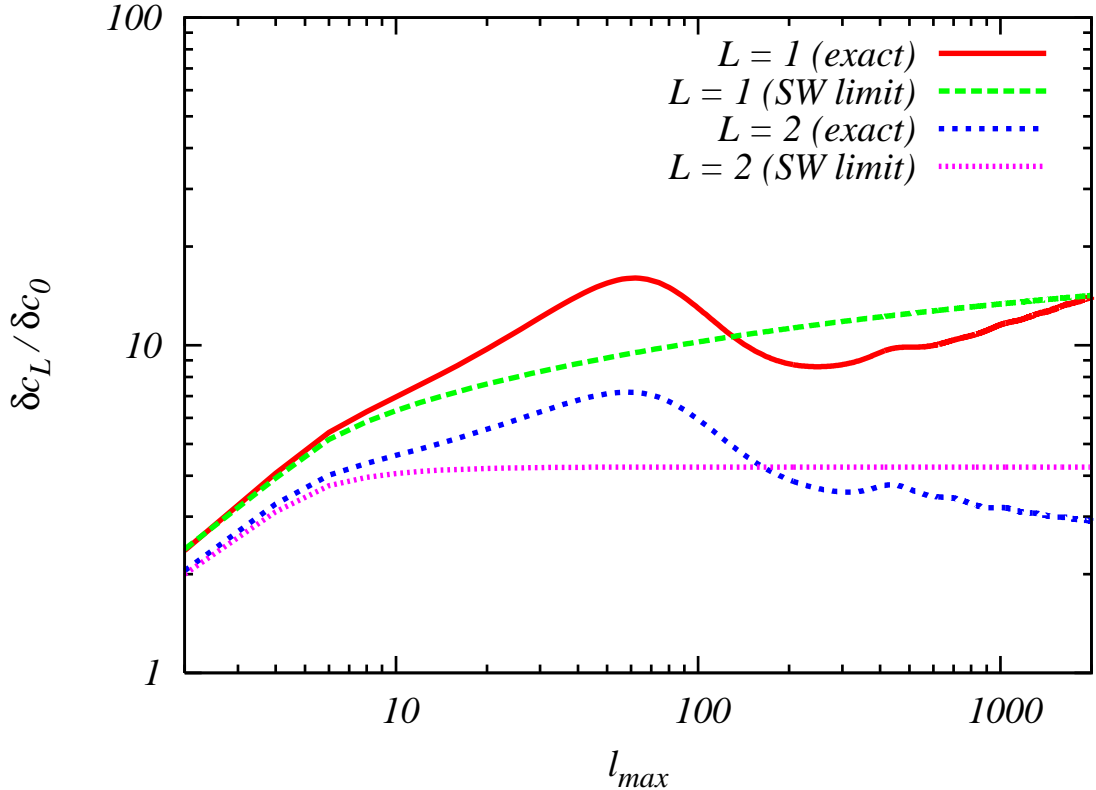}
  \end{center}
  \caption{Ratios of the expected error bars, $\delta c_L/\delta c_0$
 ($L=1$ and 2), as a function of the maximum multipoles in
 the sum, $\ell_{\rm max}$. The solid and short-dashed lines show
 the exact results for $L=1$ 
and $L=2$, respectively, while the long-dashed and dotted lines show the
corresponding Sachs--Wolfe approximations for $L=1$ and $L=2$,
respectively. We find that the Sachs--Wolfe approximations trace the
overall behavior of the exact calculations well.}
  \label{fig:variance_ratio}
\end{figure} 

In this subsection, we calculate the 1-$\sigma$ error bars of $c_0$,
$c_1$ and $c_2$, i.e., $\delta c_0$, $\delta c_1$, and $\delta c_2$,
expected for a cosmic-variance-limited experiment measuring temperature
anisotropy.  Here, we shall focus on a simultaneous estimation of a pair
of parameters: $(c_0,c_1)$ and $(c_0,c_2)$. We  give the full
constraint varying all three parameters simultaneously in
appendix~\ref{appen:3Dfit}. 

Following ref.~\cite{Komatsu:2001rj}, we calculate the Fisher matrix,
$F_{LL'}$, from 
\begin{eqnarray}
F_{L L'}\equiv \sum_{2 \leq \ell_1 \leq \ell_2 \leq \ell_3 \leq {\ell_{\rm max}}} 
\frac{B_{\ell_1 \ell_2 \ell_3}^{L}B_{\ell_1 \ell_2
\ell_3}^{L'}}{\sigma_{\ell_1 \ell_2 \ell_3}^2} ~,
\end{eqnarray}
where the variance of the CMB bispectrum, $\sigma_{\ell_1 \ell_2
\ell_3}^2$, is given by
\begin{eqnarray}
\sigma_{\ell_1 \ell_2 \ell_3}^2 = C_{\ell_1} C_{\ell_2} C_{\ell_3} 
\left[ (-1)^{\ell_1 + \ell_2 + \ell_3} 
(1 + 2 \delta_{\ell_1, \ell_2} \delta_{\ell_2, \ell_3})  
+ \delta_{\ell_1, \ell_2} + \delta_{\ell_2, \ell_3} + \delta_{\ell_3, \ell_1}
\right]~,
\end{eqnarray}
with $C_\ell$ being the power spectrum of temperature fluctuations. As
we consider a cosmic-variance-limited experiment, we ignore instrumental
noise here. 

As we show in appendix~\ref{appen:3Dfit}, $c_0$ and $c_1$ are nearly
uncorrelated, so are $c_0$ and $c_2$; however, $c_1$ and $c_2$ are
highly correlated. Therefore, in this subsection, we shall
consider submatrices of $F_{LL'}$ involving only either $(c_0,c_1)$ or
$(c_0,c_2)$, and study the full matrix in appendix~\ref{appen:3Dfit}.

We define the submatrix (a $2\times 2$ matrix) as
\begin{eqnarray}
{}^{(2)}F_{ij} \equiv \left(
\begin{array}{ccc}
  F_{00} & F_{0L} \\
  F_{L0} & F_{LL} \\ 
\end{array}
\right) ~,
\end{eqnarray}
where $L$ takes on either 1 or 2. The 1-$\sigma$ marginalized error bars
are then given by the matrix inverse as  
$(\delta c_0, \delta c_{L})  
= \left( \sqrt{ {}^{(2)}F^{-1}_{11} },  \sqrt{ {}^{(2)}F^{-1}_{22} } \right)$.

In figure~\ref{fig:variance_ratio}, we show the ratios of error bars,
$\delta c_1 / \delta c_0$ and $\delta c_2 / \delta c_0$, as a function
of the maximum multipole in the sum, $\ell_{\rm max}$. 
The solid and short-dashed lines show the exact results for $L=1$
and $L=2$, respectively, while the long-dashed and dotted lines show the
corresponding Sachs--Wolfe approximations for $L=1$ and $L=2$,
respectively. We find that the Sachs--Wolfe approximations trace the
overall behavior of the exact calculations well.

The error bar on $c_1$ is an order of magnitude larger than that on
$c_0$ for $\ell_{\rm max}\gtrsim 100$, as the $L=1$ bispectrum has a
vanishing amplitude in the squeezed limit. On the other hand, the error
bar on $c_2$ is comparable to that on $c_0$: the Sachs--Wolfe
approximation gives an asymptotic relation of $\delta c_2=4\delta
c_0$. The exact calculation gives $\delta c_2\approx 3\delta c_0$ for
$\ell_{\rm max}=2000$.

Finally, the 1-$\sigma$ error bars expected for a
cosmic-variance-limited experiment measuring temperature anisotropy up
to $\ell_{\rm max} = 2000$ are given by
\begin{eqnarray}
(\delta c_0, \delta c_1) &=& (4.4, 61) ~, \\ 
(\delta c_0, \delta c_2) &=& (4.4, 13) ~. 
\end{eqnarray}
See eq.~(\ref{eq:fisher_element}) for the full Fisher matrix.

\section{Consistency relations with higher spin fields}
\label{sec:SY}
Primordial correlation functions in the limit that some combinations of external momenta go to zero -- \emph{soft limits} -- play a special role in constraining the physics of inflation.  Significant squeezed non-Gaussianity is associated with the presence of extra light degrees of freedom during inflation, hence soft limits can be understood as probing the spectrum of light fields in the early universe.  Moreover, soft limits are observationally relevant and are subject to a number of interesting theoretical consistency relations.  The first example of such a consistency relation was noted in ref.~\cite{maldacena:2003} and established under much more general conditions in ref.~\cite{creminelli/zaldarriaga:2004}: 
$\lim_{k_3\rightarrow 0}B_\zeta(k_1,k_2,k_3)=
(1-n_s) P_\zeta(k_1) P_\zeta(k_3)$. This holds independently of the
inflationary Lagrangian, under the 
assumption that there is only a single 
field, an attractor solution has been reached \cite{namjoo/firouzjahi/sasaki:prep,chen/etal:prep}, and the
initial state is in a Bunch-Davies state
\cite{agullo/parker:2011,ganc:2011,Chialva:2011hc}.  

Our new parametrization given by eq.~(\ref{eq:bis_curv_general})
represents a non-trivial modification of 				         
this consistency relation:
\begin{equation} 
\lim_{k_3\rightarrow 0}B_\zeta(k_1,k_2,k_3)
= 
   \left( 2 \sum_L c_L P_L(\hat{\bf k}_1 \cdot \hat{\bf k}_3 )  \right) P_\zeta(k_1) P_\zeta(k_3) \, .
\end{equation}
The possibility that some of the $c_L$ coefficients can be $\gg |n_s-1|\sim 10^{-2}$ indicates extra light fields in the early universe, while the non-trivial angular dependence is associated with anisotropic sources (such as higher spin fields).  We also expect non-trivial soft limits for higher-order correlation functions.  To explore such effects, we introduce the quantities
\begin{eqnarray}
  f_{NL}^{\mathrm{eff}}(k_i) &\equiv& \lim_{k_3\rightarrow
   0} \frac{5}{12}\frac{B_\zeta(k_1,k_2,k_3)}{P_\zeta(k_1)
   P_\zeta(k_3)} \, , \label{fNL_eff} \\
 \tau_{NL}^{\mathrm{eff}}(k_i) &\equiv& \lim_{{\bf k_1}+{\bf
  k_2}\rightarrow 0} \frac{1}{4}
  \frac{T_\zeta(k_1,k_2,k_3,k_4)}{P_\zeta(|{\bf k}_1+{\bf k}_2|) P_\zeta(k_1)
  P_\zeta(k_3)} \, , 
\label{tauNL_eff}
\end{eqnarray}
where 
$\langle \zeta_{{\bf k}_1} \zeta_{{\bf k}_2} \zeta_{{\bf k}_3}\zeta_{{\bf k}_4}
\rangle=(2\pi)^3\delta^{(3)}(\sum_i{\mathbf
k}_i)T_\zeta(k_1,k_2,k_3,k_4)$.  For the local-type non-Gaussianity the quantities $f_{NL}^{\mathrm{eff}}$ and $\tau_{NL}^{\mathrm{eff}}$ become the standard (momentum independent) non-linearity parameters. In this case there is an interesting consistency relation:
\begin{equation}
\label{SY}
 \tau_{NL} \geq \left(\frac{6}{5} f_{NL}\right)^2 \, , 
\end{equation}
where we dropped the superscript ``eff'' to emphasize that this relation is understood in the case where eqs.~(\ref{fNL_eff}) and (\ref{tauNL_eff}) are independent of momenta.  This relation was first noted by Suyama and Yamaguchi in ref.~\cite{suyama/yamaguchi:2007} and further explored in refs.~\cite{komatsu:2010,sugiyama/komatsu/futamase:2011,smith/loverde/zaldarriaga:2011,Lewis:2011au,Bramante:2011zr,sugiyama:2012,assassi/baumann/green:2012}.

In this subsection, we explore the non-Gaussianity consistency relations
in the context of the $I^2(\phi)F^2$ model given in eq.~(\ref{ratra}).
In general this model breaks statistical isotropy, as discussed in
section~\ref{subsec:vectors}.  From eq.~(\ref{z3-phiF2}) we obtain
\begin{equation}
f_{NL}^{\rm eff} \simeq 10 N_{k_1}^2  N_{k_3} 
\frac{24 E_{\rm cl}^2}{\epsilon V ( \phi)}
  \left[     
1 - \cos^2 \theta_{ {\hat k}_1 ,\, {\hat E}_{\rm cl} } - \cos^2 \theta_{ {\hat k}_3 ,\, {\hat E}_{\rm cl} } 
+   \cos \theta_{ {\hat k}_1 ,\, {\hat E}_{\rm cl} } \cos \theta_{ {\hat k}_3 ,\, {\hat E}_{\rm cl} } \cos \theta_{ {\hat k}_1 ,\, {\hat k}_3 } 
\right] \, ,
\end{equation}
where we recall that $|g_*(N_{k_i})|=N_{k_i}^2\frac{24 E_{\rm
cl}^2}{\epsilon V \left( \phi \right)}  \ll  1$ is assumed.  

We compute the trispectrum for the first time in the model  given by eq.~(\ref{ratra}).  The computation follows very closely the 
analogous one of the bispectrum performed in ref.~\cite{Bartolo:2012sd}, and so we omit the technical details. We obtain  
\begin{eqnarray}
\!\!\!\! \!\!\!\!
\tau_{\rm NL}^{\rm eff} \simeq 
144 N_{k_1}^2 N_{k_3}^2 
\frac{24 E_{\rm cl}^2}{\epsilon V ( \phi)}
\Bigg[     & & \!\!\!\!
1 - \cos^2 \theta_{ {\hat k}_1 ,\, {\hat E}_{\rm cl} } -  \cos^2 \theta_{ {\hat k}_{12} ,\, {\hat E}_{\rm cl} } - 
\cos^2 \theta_{ {\hat k}_3 ,\, {\hat E}_{\rm cl} }  
\nonumber\\
& & 
+   \cos \theta_{ {\hat k}_1 ,\, {\hat E}_{\rm cl} } \cos \theta_{ {\hat
k}_3 ,\, {\hat E}_{\rm cl} } \cos \theta_{ {\hat k}_1 ,\, {\hat k}_3 }    
\nonumber\\
& & 
+   \cos \theta_{ {\hat k}_1 ,\, {\hat E}_{\rm cl} } \cos \theta_{ {\hat
k}_{12} ,\, {\hat E}_{\rm cl} } \cos \theta_{ {\hat k}_1 ,\, {\hat
k}_{12} }    
\nonumber\\
& & 
+   \cos \theta_{ {\hat k}_{12} ,\, {\hat E}_{\rm cl} } \cos \theta_{ {\hat k}_3 ,\, {\hat E}_{\rm cl} } \cos \theta_{ {\hat k}_{12} ,\, {\hat k}_3 }    
\nonumber\\
& & 
-   \cos \theta_{ {\hat k}_1 ,\, {\hat E}_{\rm cl} } \cos \theta_{ {\hat k}_3 ,\, {\hat E}_{\rm cl} } \cos \theta_{ {\hat k}_1 ,\, {\hat k}_{12} }    
\cos \theta_{ {\hat k}_{12} ,\, {\hat k}_3 }    \Bigg],
\end{eqnarray}
where ${\hat k}_{12} $ is the unit vector in the direction of ${\bf k}_1 + {\bf k}_2$.  We note that the unit-vector 
${\hat k}_{12}$ enters in this expression, even though the corresponding vector $\vec{k}_1 + \vec{k}_2$ vanishes in the squeezed limit (analogously to the  ${\hat k}_3$-dependence of the bispectrum).

We observe that, in the $I^2(\phi)F^2$ model, both $f_{NL}^{\mathrm{eff}}$ and $\tau_{NL}^{\mathrm{eff}}$ exhibit
highly non-trivial momentum dependence; thus, it is not sensible to
compare different configurations.   One can readily see that $\tau_{\rm
NL }^{\mathrm{eff}}$ vanishes for several configurations. This happens,
for example, if one of  ${\hat k}_1 \,, {\hat k}_{12} \,, {\hat k}_3$ is parallel to ${\hat E}_{\rm cl}$, while the other two vectors are perpendicular to ${\hat E}_{\rm cl}$.  We
do not interpret this as a violation of the Suyama-Yamaguchi inequality;
the expression given by eq.~(\ref{SY}) requires either that the
non-linearity parameters are momentum-independent, or else that
$\tau_{NL}^{\mathrm{eff}}$ and $\left(f_{NL}^\mathrm{eff}\right)^2$ have the same
momentum dependence so that one can factor out an amplitude which obeys
eq.~(\ref{SY}).  The original form of the Suyama-Yamaguchi inequality is
simply not applicable to the model given by
eq.~(\ref{ratra}).\footnote{ref.~\cite{Rodriguez:2013cj} presents a
detailed discussion on how to generalize the Suyama-Yamaguchi inequality
to momentum-dependent non-linearity parameters, with a particular emphasis on the case of broken statistical isotropy. The fact that we have shown that there exist some configurations with vanishing $\tau_{NL}^{\mathrm{eff}}$ implies that, at least in principle,
one may obtain observational evidence for a smaller squeezed trispectrum
than the one obtained from scalar fields, provided that one can construct an
observable quantity which is sensitive to those configurations.  It remains to be seen whether such a measurement is feasible.}

In ref.~\cite{assassi/baumann/green:2012} Assassi et al derived a general inequality relating the soft limits of the bispectrum and trispectrum:
\begin{eqnarray}
&&   \int d^3 q_1 d^3 q_2 \,\, \tau^{\mathrm{eff}}_{NL} ( {\bf q}_1, {\bf k}-{\bf q}_1, {\bf q}_2,-{\bf q}_2-{\bf k} ) \,\, P_\zeta(q_1) P_\zeta(q_2) \nonumber \\
&&   \geq \left[  \int d^3 q \,\, \frac{6}{5} \, f_{NL}^{\mathrm{eff}} ( {\bf q},   -{\bf q}-{\bf k} , {\bf k} ) \,\, P_\zeta(q) \right]^2   \, , \label{gen_ineq}
\end{eqnarray}
where the $k\rightarrow 0$ limit is understood.  This inequality reduces
to eq.~(\ref{SY}) in the local case, but is completely general and
should be respected by \emph{any} model.  We can easily verify that,
although the Suyama-Yamaguchi inequality is not meaningful here,
eq.~(\ref{gen_ineq}) is still respected.  After evaluating the angular
integral, the left hand side of eq.~(\ref{gen_ineq}) becomes
\begin{equation}
{\rm LHS} \simeq 1024 \pi^2 
\frac{24 E_{\rm cl}^2}{\epsilon V ( \phi)}
 \sin^2 \theta_{ {\hat k} ,\, {\hat E}_{\rm cl} } \left( \int 
 d q \, q^2 N_q P_\zeta \left( q \right)  \right)^2,
 \end{equation}  
and, proceeding analogously, the right hand side becomes
\begin{equation}
{\rm RHS} \simeq 1024  \pi^2 N_k^2 
\left[\frac{24 E_{\rm cl}^2}{\epsilon V ( \phi)}\right]^2
 \sin^4 \theta_{ {\hat k} ,\, {\hat E}_{\rm cl} } \left( \int 
 d q \, q^2 N_q P_\zeta \left( q \right)       \right)^2.
 \end{equation}  
Therefore, we find
\begin{equation}
\frac{\rm LHS}{\rm RHS}   \simeq \frac{1}{ \vert g_* \left( N_k \right) \vert 
 \sin^2 \theta_{ {\hat k} ,\, {\hat E}_{\rm cl} } } > 1.
 \end{equation}
Note that $|g_*| \ll 1$ has been assumed (as also required by
phenomenology) throughout this subsection. We therefore conclude that
the integrated inequality given by eq.~(\ref{gen_ineq}) is satisfied for any orientations of $\vec{k}$ relative to the classical vector field background.

\section{Conclusion}
\label{sec:conclusion}

The angle dependence of the bispectrum of primordial curvature
perturbations in the squeezed configuration is sensitive to the presence
of vector fields and non-trivial symmetry during inflation. 
In this
paper, we have explored phenomenological consequences of the new
parametrization of the bispectrum given by
eq.~(\ref{eq:bis_curv_general}): $B_\zeta= \sum_L c_L P_L(\hat{\bf k}_1
\cdot \hat{\bf 
 k}_2)P_\zeta(k_1)P_\zeta(k_2)+(\mbox{2 perm})$.
This form is physically well motivated, and
we have given three examples in section~\ref{sec:motivation}: the curvature
perturbation sourced by the anisotropic stress of magnetic fields; that
sourced by an interaction with a vector field of 
the form $I^2(\phi)F^2$; and solid inflation.

We find that a cosmic-variance-limited CMB experiment measuring
temperature anisotropy up to $\ell_{\rm max}=2000$, such as the {\sl
Planck} satellite, can measure $c_1$ and $c_2$ down to $\delta c_1=61$
and $\delta c_2=13$ (68\%~CL). The latter error bar is comparable to
(and only a factor of three larger than) the error bar of $c_0=6f_{\rm
NL}/5$; thus, if the forthcoming {\sl Planck} data reveal evidence for
$c_0$, one should also measure $c_2$ to understand the nature of sources
of non-Gaussianity. Moreover, even if the {\sl Planck} data do not reveal
evidence for $c_0$, one should still measure $c_2$, as solid inflation
can generate large $c_2$ {\it without generating detectable} $c_0$.
Sensitivity to $c_1$ is an order of magnitude worse than that to $c_0$
or $c_2$ because the term proportional to $c_1$ vanishes in the squeezed
limit due to symmetry.

The angle-dependent bispectrum in the squeezed configuration is a
natural consequence of broken statistical isotropy. Broken
isotropy also leads to a non-trivial modification of the inequality
between the local-form trispectrum amplitude, $\tau_{\rm NL}$, and
$f_{\rm NL}^2$. We find that the original form of the Suyama-Yamaguchi
inequality, $\tau_{\rm NL}\ge (6f_{\rm NL}/5)^2$,
does not apply to the current model, due to the momentum- and shape-dependence 
of $\tau_{\rm NL}$ and $f_{\rm NL}$. For example, we find some squeezed
configurations in which $\tau_{\rm NL}$ vanishes. It remains to be seen
how sensitive the forthcoming tests of the Suyama-Yamaguchi 
inequality  using the {\sl Planck} or the large-scale structure data are
to the decrease of the trispectrum amplitudes for these
particular shapes. We also find that a general inequality of
ref.~\cite{assassi/baumann/green:2012} is  satisfied in this model. 

What is next? Phenomenological consequences of
eq.~(\ref{eq:bis_curv_general}) for large-scale structure of the
universe such as the dark matter halo bias \cite{shiraishi/etal:prep},
bispectrum, and trispectrum 
should certainly be explored. For instance, a consistency relation 
\cite{Peloso:2013zw,Kehagias:2013yd} between the squeezed bispectrum and
the power spectrum of dark matter density fluctuations has been proved 
 at the full nonlinear level, and with an initial isotropic non-Gaussianity,
 in ref.~\cite{Peloso:2013zw}. It may be interesting to study a signature of 
 anisotropic initial non-Gaussianity in that context. 
Also, it is quite possible that the
coefficients $c_L$ depend on wavenumbers, which would be particularly
interesting for dipolar and quadrupolar modulations of the observed power
spectrum in our sky \cite{schmidt/hui:2013}.  Indeed, in the case of the
$I^2(\phi)F^2$ model discussed in section~\ref{subsec:vectors}, the $c_L$
coefficients exhibit a logarithmic running with wavenumber. It would be
interesting to study possible effects of such wavenumber dependence.

\section*{\it Note added}

After our paper was submitted, the Planck collaboration reported constraints on non-Gaussianity parameters \cite{planck2013fnl}. They also evaluated the coefficients $c_1$ and $c_2$ as follows: they first constrain coefficients of the basis functions of the ``modal expansion,'' from which they construct templates for the shapes corresponding to $L=1$ and 2 of eq.~(\ref{eq:bis_curv_general}). This results in a value of $11.0 \pm 113$ (68\%~CL) for the coefficient $c_1$, and  of  $3.8 \pm 27.8$ (68\%~CL) for the coefficient $c_2$. As also remarked in ref.~\cite{planck2013fnl}, the template used in their analysis is only $60\%$ correlated with $L=2$, suggesting that the estimators constructed from the modal expansion can provide an estimate for these shapes, but that they might not be optimal. Our reported forecast for the 1-$\sigma$ uncertainties, $\delta c_L$, assumes  a full sky, cosmic-variance-limited experiment measuring temperature anisotropy up to $\ell = 2000$. However, the Planck data are noise dominated for $\ell \gtrsim 1500$ and the analysis presented in ref.~\cite{planck2013fnl} uses 73\% of the sky. Rescaling our estimates by $1/\sqrt{0.73}$ but still assuming a cosmic-variance-limited experiment, we find $\delta c_0=5.1$, $\delta c_1=71$, and $\delta c_2=15$. On the other hand, the Planck collaboration finds $\delta c_0=7.0$, $\delta c_1=113$, and $\delta c_2=28$ (recall that $c_0$ is equal to $6f_{\rm NL}/5$). As the Planck collaboration uses the optimal estimator to find a limit on $\delta c_0$, we estimate the effect of noise in the Planck data by rescaling the error bars by the ratio of $7.0/5.1$. We find $\delta c_1=97$ and $\delta c_2=21$. These estimates for $\delta c_1$ and $\delta c_2$ are $16\%$ and $33\%$ lower than the error bars that the Planck collaboration finds. The latter can be understood from the fact that the template for $L=2$ used by the Planck collaboration is $60\%$ correlated with the true shape. Therefore, it appears that there is still some room for improvement in the limits on these parameters, especially $c_2$, using optimal estimators.

\acknowledgments
MP and EK thank Kavli Institute for the Physics and Mathematics of the
Universe (Kavli IPMU, WPI), where this work was initiated, for
hospitality during our stay. MS and EK thank the organizers of the
Long-term Workshop YITP-T-12-03 on ``Gravity and Cosmology 2012'' held at the
Yukawa Institute for Theoretical Physics, Kyoto University, during which
the initial draft of this paper was written. This work was supported in
part by a Grant-in-Aid for JSPS Research under Grant No.~22-7477 (MS),
and Grant-in-Aid for Nagoya University Global COE Program ``Quest for
Fundamental Principles in the Universe: from Particles to the Solar
System and the Cosmos,'' from the Ministry of Education, Culture,
Sports, Science and Technology of Japan. We also acknowledge the
Kobayashi-Maskawa Institute for the Origin of Particles and the
Universe, Nagoya University, for providing computing resources. 
MP would like to thank the University of Padova, and INFN, Sezione di
Padova, for their friendly hospitality and for partial support during
his sabbatical leave. MP would also like to thank Nicola Bartolo, Antony Lewis, and Sabino Matarrese for valuable discussions. We also thank Nicola Bartolo, Michele Liguori, and Sabino Matarrese for discussion about the PLANCK results. NB is grateful to Valentin Assassi and Daniel Baumann for interesting discussions. We thank Christian Byrnes for useful correspondence. NB and MP acknowledge partial support from the DOE grant DE-FG02-94ER-40823 at the University of Minnesota.   

\appendix
\section{Precision of the flat-sky approximation}\label{app:flat}
\begin{figure}[t]
  \begin{tabular}{c}
    \begin{minipage}{1\hsize}
  \begin{center}
    \includegraphics[width = 0.85\textwidth]{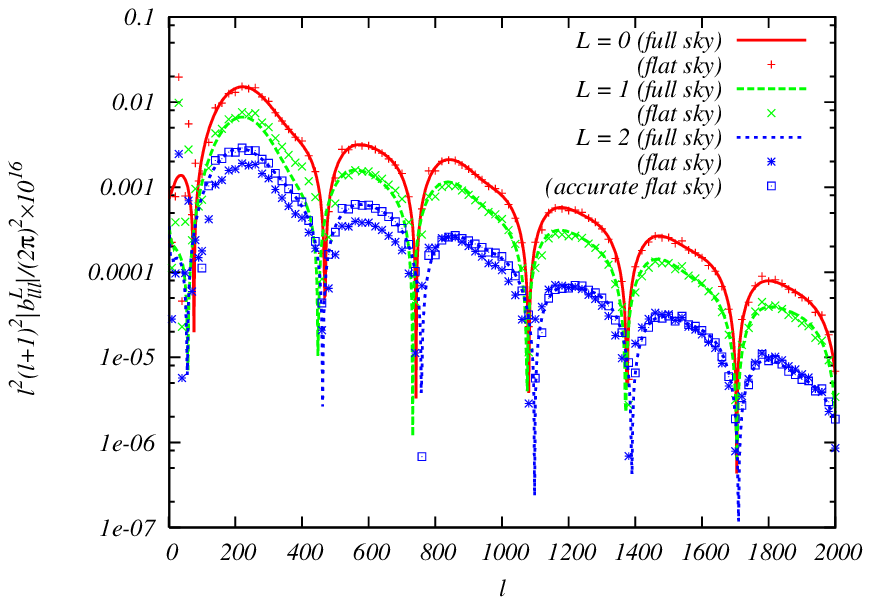}
  \end{center}
\end{minipage}
  \end{tabular}
  \begin{tabular}{c}
\begin{minipage}{1\hsize}
  \begin{center}
    \includegraphics[width = 0.85\textwidth]{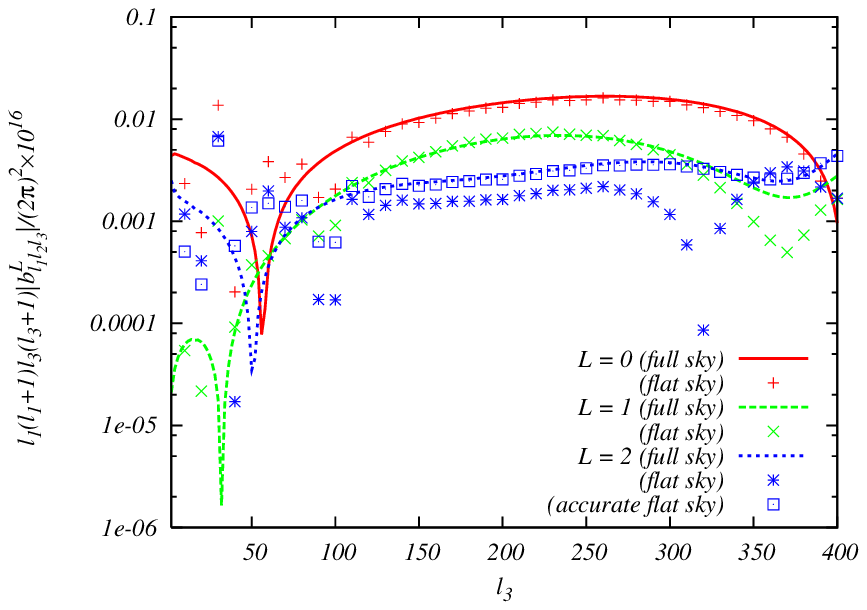}
  \end{center}
\end{minipage}
\end{tabular}
  \caption{Absolute values of the CMB temperature reduced bispectra. The
 solid, long-dashed and short-dashed lines show the full-sky results for
 $L=0$, 1, and 2, respectively, while the plus, cross, and star symbols
 show the simplified flat-sky results from
 eq.~(\ref{eq:flatskybispectrum}) for $L=0$, 1, and 2, respectively. The
 square symbols show the original form of the flat-sky result for $L=2$ from
 eq.~(\ref{eq:originalflat}) before further approximation.
(Top panel) Equilateral triangles, $|b_{\ell\ell\ell}^L|$. 
(Bottom panel) Squeezed triangles, $|b_{\ell_1\ell_2\ell_3}^L|$, with $\ell_1 =
 \ell_2 = 200$, as a function of $\ell_3$.}
\label{fig:flatcheck}
\end{figure}

How precise are the flat-sky formulae given by eqs.~(\ref{eq:originalflat})
and (\ref{eq:flatskybispectrum})? In figure~\ref{fig:flatcheck}, we
compare the full-sky results with the flat-sky results. We find that,
for $L=0$ and 1, the simplified flat-sky formula given by
eq.~(\ref{eq:flatskybispectrum}) yields the bispectra in the equilateral
and squeezed configurations  which are in good agreement with the
full-sky results at $\ell\gtrsim 100$. However, we find that, for $L=2$,
the simplified 
formula systematically underestimate the magnitude of the bispectra in
both configurations. The equilateral result suggests that the simplified
formula provides an adequate result only at $\ell\gtrsim 800$.

While these results appear to suggest that the precision of the
simplified formula degrades as $L$ increases, this is not the case: the
flat-sky results in the Sachs-Wolfe limit (which are not shown in
this paper) show that the simplified formula {\it overestimates}
the magnitudes of the bispectra of $L=1$ and 2 in the Sachs--Wolfe limit
by a similar amount in both equilateral and squeezed 
configurations. Therefore, we conclude that the
simplified formula given by eq.~(\ref{eq:flatskybispectrum}) should only
be used for quantitative calculations of $L=0$ or for {\it
qualitative} calculations of $L=1$ and 2, and the original formula given by
eq.~(\ref{eq:originalflat}) should be used for
quantitative calculations of $L=1$ and 2. Needless to say, the full-sky
formula should always be used for the calculations involving multipoles
of $\ell\lesssim 100$.

\section{Analysis in the Sachs-Wolfe limit}\label{appen:SW}
As the calculations presented in section~\ref{sec:full} are quite involved,
some appropriate approximations would be useful for understanding the
analytical structures of the basic results. 

The Sachs--Wolfe limit, in which the radiation transfer function
is given by ${\cal T}_{\ell}(k) \to - \frac{1}{5} j_\ell (k r_*)$,
provides such a convenient approximation. With this transfer function,
$\alpha_\ell(r)$ (eq.~\ref{eq:alpha}) simplifies to 
$\alpha_{\ell}(r) \to - \frac{1}{5 r_*^2} \delta(r - r_*)$,
where $r_* \equiv \tau_0 -\tau_*$ is the conformal distance to the last
scattering surface. Similarly, for a scale-invariant spectrum of
$\zeta$, $P_\zeta(k)=\frac{2\pi^2}{k^3}A_S$, $\beta_\ell(r)$ (eq.~\ref{eq:beta}) becomes 
\begin{eqnarray}
\beta_{\ell L}(r_*) \to - \frac{\pi^2}{10} A_S 
\frac{\Gamma\left(\frac{\ell + L}{2}\right)}{\Gamma\left(\frac{\ell - L + 3}{2}\right) \Gamma\left(\frac{-\ell + L + 3}{2}\right) 
\Gamma\left(\frac{\ell + L + 4}{2}\right)} ~,
\end{eqnarray} 
where $\Gamma(x)$ is the Gamma function.
Using these $\alpha_\ell$ and $\beta_\ell$ in
eq.~(\ref{eq:CMB_bis_comp}), one finds the Sachs--Wolfe approximation of
the CMB bispectrum as 
 \begin{eqnarray}
 B_{\ell_1 \ell_2 \ell_3}^{L} &\to&  
- \frac{1}{5} 
\left[ \prod_{n=1}^3 \sum_{L_n} 
 (-1)^{\frac{\ell_n + L_n}{2}}  \right] 
I_{L_1 L_2 L_3}
\beta_{\ell_1 L_1}(r_*) \beta_{\ell_2 L_2}(r_*) 
 \nonumber \\ 
&&\times 
\frac{4\pi}{2L+1}  
I_{\ell_1 L_1 L}
 I_{\ell_2 L_2 L}
(-1)^{\ell_2 + L_1} \delta_{L_3, \ell_3}
\left\{
  \begin{array}{ccc}
  \ell_1 & \ell_2 & \ell_3 \\
  L_2 & L_1 & L 
  \end{array}
 \right\}
+ (\mbox{2 perm}). \label{eq:cmb_bis_legendre_SW}
\end{eqnarray}

\begin{figure}[t]
  \begin{tabular}{cc}
    \begin{minipage}{0.5\hsize}
  \begin{center}
    \includegraphics[width = 7.5 cm]{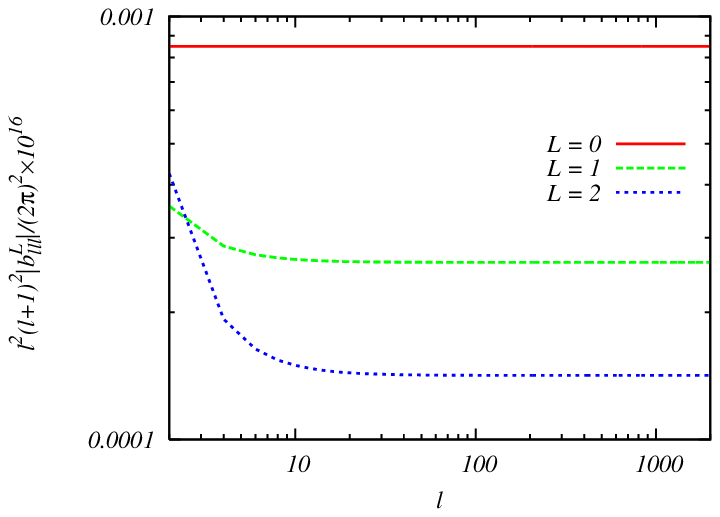}
  \end{center}
\end{minipage}
\begin{minipage}{0.5\hsize}
  \begin{center}
    \includegraphics[width = 7.5 cm]{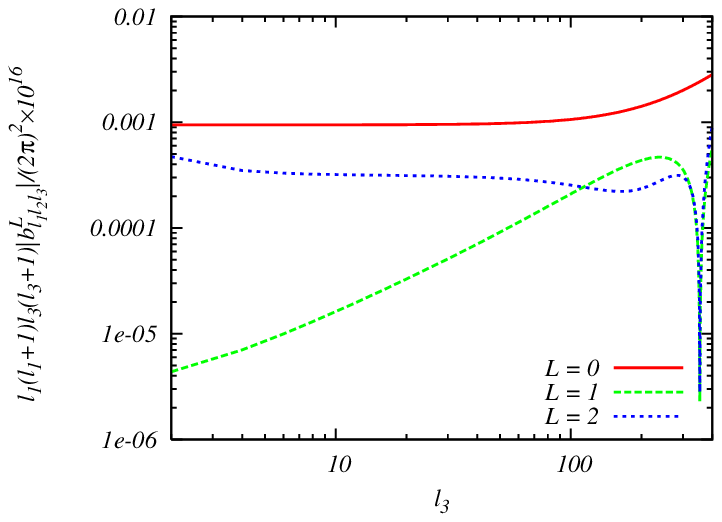}
  \end{center}
\end{minipage}
\end{tabular}
  \caption{Absolute values of the CMB temperature reduced bispectra for
 $L=0$ (solid), 1 (long-dashed), and 2 (short-dashed), in the
 Sachs--Wolfe limit. 
(Left panel) Equilateral triangles, $|b_{\ell\ell\ell}^L|$. (Right
 panel) Squeezed triangles, $|b_{\ell_1\ell_2\ell_3}^L|$, with $\ell_1 =
 \ell_2 = 200$, as a function of $\ell_3$.}
\label{fig:SSS_III_SW}
\end{figure}

Figure~\ref{fig:SSS_III_SW} shows the reduced CMB temperature bispectra
in the Sachs--Wolfe limit. The basic behaviors, such as the monotonic
decrease of the equilateral amplitudes as a function of $L$ and the
suppression of the $L=1$ term in the squeezed limit, are all reproduced
by the simple Sachs--Wolfe limit calculations. These results may be
compared with figures~\ref{fig:SSS_equ} and \ref{fig:SSS_squ}. 

\section{Full Fisher matrix}\label{appen:3Dfit}

In section~\ref{sec:2DFisher}, we have presented the 1-$\sigma$
marginalized constraints on $c_0$, $c_1$, and $c_2$, assuming that only
$c_0$ and one of $c_1$ and $c_2$ are varied simultaneously. In this
appendix, we provide the full Fisher matrix, $F_{LL'}$, involving all of
$c_0$, $c_1$, and $c_2$, calculated up to $\ell_{\rm max}=2000$:
\begin{eqnarray}
F_{LL'}= \left(
\begin{array}{ccc}
  5232 & 16.94 & -5.986 \\
  16.94 & 26.53 & 66.85 \\ 
  -5.986 & 66.85 & 618.1 
\end{array}
\right) \times 10^{-5}
~. \label{eq:fisher_element}
\end{eqnarray}

From this matrix, one can compute the cross-correlation coefficients,
$r_{LL'}\equiv F_{LL'}/\sqrt{F_{LL}F_{L'L'}}$. We find 
$r_{01}=0.045$ and $r_{02}=-0.003$, indicating that $c_0$ and $c_1$
are nearly uncorrelated, so are $c_0$ and $c_2$. However, there is a
high degree of correlation between $c_1$ and $c_2$: $r_{12}=0.522$.
As a result, the marginalized error bars increase slightly to  
\begin{eqnarray}
(\delta c_0, \delta c_1, \delta c_2)
= \left( \sqrt{ F^{-1}_{00} }, 
\sqrt{ F^{-1}_{11} }, 
\sqrt{ F^{-1}_{22} } \right) 
= (4.4, 72, 15) ~.
\end{eqnarray}

\bibliography{references}
\end{document}